\documentclass[journal,draftcls,onecolumn,12pt,twoside]{IEEEtran}
\IEEEoverridecommandlockouts
\usepackage{cite}
\usepackage{tabularx}
\usepackage{amsmath,amssymb,amsfonts}
\usepackage{algorithmic}
\usepackage{multirow}
\usepackage{ragged2e,array,booktabs}
\usepackage{graphicx}
\usepackage{textcomp}
\usepackage{xcolor}
\usepackage{hyperref}
\usepackage{bookmark}
\usepackage{subfigure}
\usepackage{float} 
\def\BibTeX{{\rm B\kern-.05em{\sc i\kern-.025em b}\kern-.08em
		T\kern-.1667em\lower.7ex\hbox{E}\kern-.125emX}}
 
\begin{document}
	
\title{Drug Release Management for Dynamic TDMA-Based  Molecular Communication}
\author{Hamid Khoshfekr Rudsari, Nader Mokari,~\IEEEmembership{Senior Member,~IEEE}, Mohammad Reza Javan,~\IEEEmembership{Member,~IEEE}, Eduard A. Jorswieck,~\IEEEmembership{Senior Member,~IEEE}, and Mahdi Orooji,~\IEEEmembership{Member,~IEEE}
	\thanks{Hamid Khoshfekr Rudsari, Nader Mokari and Mahdi Orooji are with the Department of Electrical and Computer Engineering, Tarbiat Modares University, Tehran, Iran. e-mail: \{hamid\_khoshfekr, nader.mokari, morooji\}@modares.ac.ir}	
	\thanks{Mohammad Reza Javan is with the Department of Electrical and Robotic Engineering, Shahrood University of Technology, Shahrood, Iran. e-mail: javan1378@yahoo.com}
	\thanks{E. A. Jorswieck is with the Department of Information Theory and Communication Systems, TU Braunschweig, Braunschweig, Germany. e-mail: jorswieck@ifn.ing.tu-bs.de}}
\maketitle
\markboth{IEEE Transactions on Communications}%
{}
\begin{abstract} 
	In this paper, we design a drug release mechanism for dynamic time division multiple access (TDMA)-based molecular communication via diffusion (MCvD). In the proposed scheme, the communication frame is divided into several time slots over each of which a transmitter nanomachine is scheduled to convey its information by releasing the molecules into the medium. To optimize the number of released molecules and the time duration of each time slot (symbol duration), we formulate a multi-objective optimization problem whose objective functions are the bit error rate (BER) of each transmitter nanomachine. Based on the number of released molecules and symbol durations, we consider four cases, namely: ``static-time static-number of molecules'' (STSN), ``static-time dynamic-number of molecules'' (STDN), ``dynamic-time static-number of molecules'' (DTSN), and ``dynamic-time dynamic-number of molecules'' (DTDN).  We consider three types of medium in which the molecules are propagated, namely: ``mild diffusive environment'' (MDE), ``moderate diffusive environment'' (MODE), and ``severe diffusive environment'' (SDE). For the channel model, we consider a 3-dimensional (3D) diffusive environment, such as blood, with drift in three directions. Simulation results show that the STSN approach is the least complex one with BER around $\text{10}^{\text{-2}}$, but, the DTDN is the most complex scenario with the BER around $\text{10}^{\text{-8}}$.
\end{abstract}

\section{Introduction} \label{sec:introduction}

\subsection{State of the Art} \label{subse:state_of_the_art}
The demand for more effective and less invasive health care solutions has pushed the technology to progress in micro and nano-scale paradigms. One of these paradigms is nano-networks, which contain nano-transmitters and nano-receivers. They communicate with each other via Molecular Communication (MC) where the molecule propagation is the main paradigm. The transmitters and receivers in MC, called bio-nanomachine, are made of biological materials and mechanisms  which are able to interact with biological molecules and cells \cite{nakano2012molecular}. The bio-nanomachines are purified protein molecules and bio-silicon hybrid devices~\cite{liu2012principles,doktycz2007nano,moore2006design}. They communicate with each other by releasing molecules into the medium. The most common medium in MC is the blood vessel. 

The main reason of introducing the Time Division Multiple Access (TDMA) technique is its proficiency in the drug management for the novel Drug Delivery System (DDS). In such a novel DDS, the management of the drug released into the blood vessels or other tissues plays a significant role. In this regard, the aim is to optimize the drug dosage and time of releasing the molecules into the medium towards the location of diseases, e.g., cancer cells. Despite the importance of the accurate releasing time and releasing dosage in the DDS, drug management is not considered in recent works while the investigation of such novel DDSs can significantly improve the performance of these systems.

\subsection{Related Works}\label{subsec:related_works}
Researchers study the TDMA optimization in neuron-based MC, which employs neurons to communicate and built in-body sensor-actuator networks (IBSANs)~\cite{suzuki2012multiobjective}. They use an evolutionary multi-objective optimization algorithm to design the TDMA schedule. The resource allocation in MC has already studied for two transmitter nodes in~\cite{jiang2015nanoscale} where the authors propose a game-theoretic framework and study Bit Error Rate (BER) of such a system. In addition, the investigation of the channel capacity for multiple-access channels, which employs the principles of natural ligand-receptor is studied in~\cite{atakan2009single}. Furthermore, the researchers have found a high capacity in Single-input Single-output (SISO) and Multi-input Single-output (MISO)-based MC system~\cite{atakan2009single}.

The investigation of more than two transmitter nodes in multiple access channel in existing works has not been considered yet. In addition, TDMA in Molecular Communication via Diffusion (MCvD) system has not been studied in the existing works on MC. The optimization of symbol durations and the number of released molecules by each transmitter node is also not considered in the existing works.

\subsection{Our Contributions} \label{subsec:contributions}
In this paper, we investigate the TDMA-based MCvD system for drug releasing management which is applicable in novel DDSs. First of all, we employ the Brownian motion~\cite{mori1965transport} as the model of propagating the molecules into the medium which is assumed to be like the blood vessels. We consider a 3-dimensional (3D) diffusive environment where the drift is assumed to exists in all the three dimensions. We assume that the channel is shared between transmitters based on TDMA method where the communication frame is divided into time slots each of which is dedicated to a transmitter. We consider Inter-user Interference (IUI) and Inter-symbol Interference (ISI) in this paper. We formulate a multi-objective optimization problem in which we determine the number of released molecules by each transmitter as well as the time duration of each time slot such that the BER of each transmitter node is minimized. We derive the mean and the variance of the number of received molecules from each transmitter node in each time slot as well as the BER. We consider four cases in each of which the number of released molecules and the duration of time slots could be fixed or dynamic obtained via the optimization problem. We first introduce the ``Static-Time Static-Number of molecules'' (STSN) in which the number of molecules and time slot durations are uniformly allocated to each transmitter. Next, the ``Dynamic-Time Static-Number of molecules'' (DTSN) is introduced in which the time slot durations are optimized via the the optimization problem while the number of molecules are uniformly allocated to each transmitter. After that, the ``Static-Time Dynamic-Number of molecules'' (STDN) is proposed in which we determine the optimized number of molecules each transmitter should release into the medium. The last case is ``Dynamic-Time Dynamic-Number of molecules'' (DTDN) which optimizes both the time slot duration and number of molecules released by each transmitter.

In addition, we consider three scenarios for the diffusion of the medium into which the molecules are propagated: 1: Mild Diffusive Environment (MDE) in which the molecules diffuse more slowly in the medium, 2: MOderate Diffusive Environment (MODE) where the molecules diffuse faster than MDE scenario, 3: Severe Diffusive Environment (SDE) where the molecules diffuse faster compared to previous two scenarios. We compare the minimum achievable BER of each three aforementioned scenarios.

We also provide the Multi-objective OPtimization~(MOP) solution as the Weighted Sum Method (WSM) technique to transform the multi-objective optimization problems into the single objective optimization problem. The main contributions of this paper are:
  \begin{itemize}
\item We consider four cases, i.e., STSN, STDN, DTSN, and DTDN, for managing the drug releasing mechanism.
\item We study the mean and the variance of the received molecules by considering the effect of the interference. Furthermore, we derive the BER for each transmitter nodes.
\item We formulate optimization problems aiming at finding the optimized time slot durations and the number of molecules released by each transmitter node considering both of dynamic and static behavior of the system.
\item We investigate the TDMA-based MCvD system in three scenarios for the the propagation medium, namely MDE, MODE, and SDE.
  \end{itemize}
 
 The remainder of the paper is organized as follows. In Section~\ref{sec:system_model}, we study the mean and the variance of the received molecules and derive the BER of each transmitter node in TDMA-based MCvD system. In Section~\ref{sec:TDMA}, we formulate the optimization problems which include the dynamic and the static cases of the time slot durations and the number of released molecules to manage the drug release mechanism. In Section~\ref{sec:multi_objective_optimization}, we provide the solution of the MOP by employing WSM. In Section~\ref{sec:comp_complexity}, we calculate the computational complexity of the proposed algorithms. We demonstrate the numerical analysis of the introduced TDMA-based molecular communication system in Section~\ref{sec:numerical_analysis}. Finally, we conclude the paper in Section~\ref{sec:conclusion}.

  \textit{Notation}: In this paper, $\exp(x)$ denotes the natural exponential function. $B(n,p)$ and $\mathcal{N}(\mu,\sigma^2)$ refer to the Binomial distribution with parameters $n$ and $p$, and the Normal distribution with mean $\mu$  and variance $\sigma^2$, respectively. Pr($X$) is the probability of happening event $X$, and $ \text{erf} \big(y\big) = \frac{1}{\sqrt{2\pi}} \int_{0}^{y}\exp\big(-x^2\big)\, dx $ stands for the error function. The abbreviations used in this paper is listed in Table~\ref{table_abbreviation}.
  \begin{table}[]
  	\caption{The list of abbreviations}
  	\centering
  	\tiny
  	\begin{tabular}{|l|l|l|l|} 
  		\hline
  		Abbreviation& Explanation & Abbreviation& Explanation \\ \hline \hline
  		ASM&Alternative Search Method&MDE&Mild Diffusive Environment\\ \hline
  		BER&Bit Error Rate&MISO&Multi-Input Single-Output\\ \hline
  		CCP&Computational Complexity Potency&  ML& Maximum Likelihood\\ \hline
  		CCS& Cartesian Coordinate System& MODE&MOderate Diffusive Environment\\ \hline
  		CDF&Cumulative Distribution Function& MOP&Multi-objective OPtimization\\ \hline
  		DDS&Drug Delivery System&OOK&On-Off Keying\\ \hline
  		DTDN&Dynamic-Time Dynamic-Number of molecules& PDF&Probability Density Function\\ \hline
  		DTSN&Dynamic-Time Static-Number of molecules& ROC&Rank Order Centroid \\ \hline
  		HIV&Human Immunodeficiency Virus&SDE&Severe Diffusive Environment\\ \hline
  		IBSAN&In-Body Sensor-Actuator Network&SISO&Single-Input Single-Output\\ \hline
  		ISI&Inter-Symbol Interference&STDN&Static-Time Dynamic-Number of molecules\\ \hline
  		IUI&Inter-User Interference&STSN&Static-Time Static-Number of molecules\\ \hline
  		MC&Molecular Communication&TDMA&Time Division Multiple Access\\ \hline
  		MCvD&Molecular Communication via Diffusion&WSM&Weighted Sum Method\\ \hline
  	\end{tabular}
  	\label{table_abbreviation}
  \end{table}
\section{System Model} \label{sec:system_model}
In this paper, we assume a multiple-access MCvD system consisting of $r$ transmitter nanomachines (nodes TX-I, TX-II, TX-III, ..., TX-$r$) and a receiver nanomachine (node D). The transmitter nanomachines are considered as the generic nano-transmitters which are produced artificially. The pre-encoded nano-transmitters are the other cases of nano-transmitters, but they are not expandable after transmitting all the stored information molecules ~\cite{chude2019nanosystems}, and therefore, we do not utilize these models of nano-transmitters. Furthermore, the receiver is a passive spherical nanomachine counting the received molecules. This type of receiver can absorb no molecules, and hence, they are not disappeared from the medium~\cite{noel2016active,noel2014improving}. It is also assumed that the distance between each transmitter node and node D can be different. In our system model, we assume that the transmitter and receiver nanomachines are fixed during the transmission period. This assumption is practical in all DDS scenarios that the nanomachines are stuck in the blood vessels such as the cases that the cancer cells are the target of the drug delivery~\cite{femminella2015molecular}. In this paper, for the medium, we consider 3D diffusive environment where the blood is an example of such a medium. In addition, the medium has drift velocity in all the three directions.
 
\begin{figure}[t]
 	\centering
 	\scalebox{.1}{}
 	\includegraphics[width=300pt]{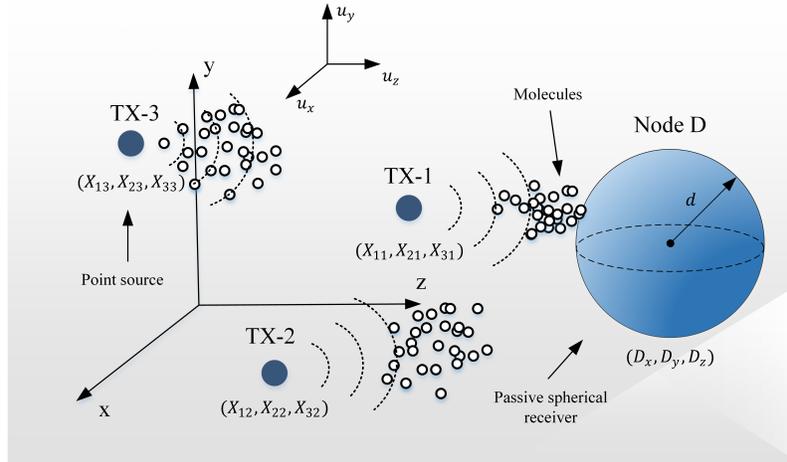}
 	\caption{Dynamic TDMA MCvD system with 3 transmitters in one frame ($r=3$). }
 	\label{Fig:dynamic_TDMA}
 \end{figure}

 Flow in diffusive environments is categorized in two classes: laminar and turbulent. If the flow stochastically varies over the time and/or space, it is categorized as turbulent, and otherwise as laminar~\cite{finnemore2002fluid}.
 With the flow in the effective length $d_{\text{eff}}$ and the effective velocity $v_{\text{eff}}$, the Reynolds number is used to determine the class of the medium, which is given by~\cite{finnemore2002fluid} 
 \begin{align}
 	Re = \dfrac{d_{\text{eff}}  \ v_{\text{eff}}}{\nu}, \label{eq:reynolds}
 \end{align}
 where $\nu$ is the kinematic viscosity of the fluid. The fluid with $Re \gg $2100 is categorized as turbulent, and otherwise in laminar. We emphasis that lots of the blood vessels have the Reynolds number smaller than 500~\cite{finnemore2002fluid}. However, both cases of laminar and turbulent are covered well in the channel model introduced in~\cite{bhatnagar20193}. The proposed MCvD scheme in our system model is demonstrated in Fig.~\ref{Fig:dynamic_TDMA}. In this figure, we assume three transmitter nodes as TX-1, TX-2, and TX-3, and one receiver node as node D. The transmitters are located at different locations and the medium is considered as a 3D unbounded diffusive environment with drift.

 We introduce 3 scenarios for the medium that molecules propagated into. The scenarios are based on the diffusion coefficient of the medium which is derived as follows~\cite{tyrrell2013diffusion}:
 \begin{align}
 	\Omega = \dfrac{k_B \Delta}{6 \pi \eta R_s}, \label{eq:diffusion}
 \end{align}
 where $k_B$, $\Delta$, $\eta$, and $R_s$ are the Boltzmann's constant~\cite{hoover1986constant}, the temperature of the medium in Kelvin~\cite{stimson1955heat}, the dynamic viscosity of the medium, and the Stoke's radius of the molecules propagated into the medium, respectively. The first scenario is MDE, where the diffusion coefficient is low. The BER of the system in this scenario has less performance compared to other ones. It is due to the fact that the molecules diffuse slowly in the considered environment. The second scenario is MODE, where the diffusion coefficient of the medium is more than that of the MDE scenario. In this case, the BER of the system has better performance compared to the MDE one. The third scenario is SDE, where the diffusion coefficient of the medium is more than that of MDE and MODE. In this scenario, the BER of the system is the best among the other scenarios because the molecules diffuse faster.

\subsection{Channel Model} \label{subsec:channel_model}
In this paper, we consider the Brownian motion to model the propagation of the molecules into the medium and a 3D diffusive medium with drift. As studied in the literature~\cite{bhatnagar20193}, the probability density function (PDF) of the molecules released from the origin in Cartesian Coordinate System (CCS) and arrived at the location of $(x,y,z)$ in CCS within time $t$ is 
\begin{align}
\upsilon(x,y,z,t) = \dfrac{1}{\sqrt{(4 \pi \Omega t)^3}} \exp \bigg( -\dfrac{(x + D_x - u_x t)^2 + (y + D_y - u_y t)^2 + (z + D_z - u_z t)^2 }{4 \Omega  t}  \bigg), \label{eq:PDF}
\end{align}
where $\Omega$, $\textbf{D} = (D_x, D_y, D_z)$, and $\textbf{u} = (u_x, u_y, u_z)$ are the diffusion coefficient of the medium,  the location of the receiver node, and the vector of drift velocity, respectively. To obtain the Cumulative Distribution Function (CDF) of the molecules inside the spherical receiver with radius $d$, we should integrate (\ref{eq:PDF}) over the volume of the receiver as follows:
\begin{align}
\begin{split}
\Upsilon (t) =& \int_{-d}^{d} \int_{-\sqrt{d^2 - z^2}}^{\sqrt{d^2 - z^2}} \int_{-\sqrt{d^2 - z^2 - y^2}}^{\sqrt{d^2 - z^2 - y^2}} \dfrac{1}{\sqrt{(4 \pi \Omega t)^3}} \\& \times \exp \bigg( -\dfrac{(z + D_z - u_z t)^2 + (y + D_y - u_y t)^2 + (x + D_x - u_x t)^2}{4 \Omega t} \bigg)~\text{d}x~\text{d}y~\text{d}z. \label{eq:CDF}
\end{split}
\end{align} 
Since (\ref{eq:CDF}) does not have a closed form solution, we can consider it as a function of the location of the destination node ($\textbf{D}$), diffusion coefficient ($\Omega$), drift velocity of the medium ($\textbf{u}$), and time ($t$). Therefore, a new notation, i.e.,  $P_{\text{arv}}(\textbf{u},\Omega,\textbf{D},t) = \Upsilon(t)$, is used in the rest of the paper. It is worth noting that by locating the transmitter node TX-$s$ at $\mathbf{X_s}$ = ($X_{1s}$, $X_{2s}$, $X_{3s}$), the probability of arriving the molecules within time slot $t_s$ is changed to $P_{\text{arv}}(\textbf{u},\Omega,\textbf{D} - \mathbf{X_s},t_s)$~\cite{rudsari2019non}. It is also concluded from (\ref{eq:CDF}) that the probability of reception of the molecules in MDE is lower than that of MODE and SDE, which in Section~\ref{sec:numerical_analysis}, we discuss the aforementioned scenarios in detail.

The activation of the transmitter nodes to release the molecules into the medium is based on the closeness of them to node D, i.e., if TX-1 is closer than TX-2 to node D, it implies TX-1 is activated first. This results in the reduction of the IUI and ISI effects. We assume the transmitter nodes release the molecules independently into the medium. Each frame is divided into $r$ slots of different lengths. Each slot has one symbol and the time slot is called the symbol duration. Therefore, the transmitters send $r$ symbols in each frame. The symbol durations are calculated by solving the optimization problems, which are introduced in Section~\ref{sec:TDMA}. Hence, the symbol durations are not identical and can be different for each transmitter.
 
To release the molecules from transmitter nodes, we adopt the one-off keying (OOK) modulation which is efficient in case of the molecules reception probability~\cite{garralda2011diffusion}. In OOK, transmitter node TX-$s$ releases $A_s$ molecules to send bit ``1''  at the beginning of a time slot, and zero molecules to send bit ``0''. We do not consume the same number of molecules for all the transmitter nodes \footnote{We use just one type of molecules in this paper, but for reducing the interference effect one can use different types of molecules for each transmitter node as a future work.}. Let us denote the number of molecules that is counted by the receiver from TX-$s$ at the $n^{\text{th}}$ frame  by $M^s[n]$. Since the movement of molecules are independent of each other, $M^s[n]$ obeys a binomial distribution as follows~\cite{singhal2015performance,kim2014symbol,kuran2012interference}:
\begin{align}
M^s[n] \sim B\bigg(x_s[n]A_s,P_h(\mathbf{X_s},t_s)\bigg), \label{eq:current_binomial}
\end{align}
where $x_s[n]$ is the information bit transmitted by node TX-$s$ at the $n^{\text{th}}$ frame and $P_h(\mathbf{X_s},t_s) =  P_{\text{arv}}(\textbf{u},\Omega,\textbf{D} - \mathbf{X_s},t_s)$.

Due to the cumbersome mathematical manipulations with binomial distribution, we aim at approximating it to the normal distribution. If $A_s$ is large enough and $A_s P_h(\mathbf{X_s},t_s)$ is not zero~\cite{papoulis2002probability}, we can approximate the binomial distribution in (\ref{eq:current_binomial}) by a normal distribution as follows:
\begin{align}
\begin{split}
M^s[n] \sim \ \mathcal{N}\bigg(x_s[n]A_s P_h(\mathbf{X_s},t_s) \ , \ x_s[n]A_s P_h(\mathbf{X_s},t_s)  (1 - P_h(\mathbf{X_s},t_s)) \bigg). \label{eq:current_normal}
\end{split}
\end{align}

\subsection{Interference Analysis} \label{subsec:ISI_IUI}
In this subsection, we study the effect of the interference on the multiple-access MCvD system. In MC systems, the interference is defined by the molecules that are leaked to the current time slot but are transmitted from previous time slots~\cite{yilmaz2015arrival,kim2013novel}.

Let us assume the current frame and transmitter as the $n^{\text{th}}$ and TX-$s$, respectively. We aim to find the distribution of molecules that is leaked into the current time slot but released from previous time slots which could be from the current frame and the previous frames. Therefore, the interference is calculated as the molecules released from the current transmitter (TX-$s$) in previous frames which is referred to ISI and the molecules released from other transmitters at the current frame and the previous frames which is referred to IUI. To be straightforward, we refer all the considered interferences (ISI and IUI) as IUI. As stated in the literature~\cite{jiang2015nanoscale,tepekule2015isi,kilinc2013receiver,noel2014unifying}, the IUI length, i.e., the number of previous frames which are involved in IUI, is limited, and therefore, we assess $U$ previous frames to calculate the IUI effect. 
By considering the aforementioned explanation, $M^s_{\text{IUI}} [n]$ which is the number of molecules in IUI, is calculated for the given TX-$s$ from $U$ previous frames as follows:
\begin{align}
\begin{split}
	M^s_{\text{IUI}} [n] \sim&   \sum_{u = 1}^{U} \sum_{j=1}^{r} B\bigg(A_j \ x_j[n - u] ,  Y^u_{j,s}\bigg) + \sum_{j=1}^{s-1} B\bigg(A_j \ x_j[n] ,  H_{j,s}\bigg), \label{eq:IUI}
\end{split}
\end{align}
where $Y^u_{j,s} = P_h(\mathbf{X_j},\lambda^u_{j,s}) - P_h(\mathbf{X_s}, \lambda^u_{j,s} - t_s)$ and $H_{j,s} = P_h(\mathbf{X_j},\sum_{q=j}^{s} t_q) - P_h(\mathbf{X_s}, \sum_{q=j}^{s-1}t_q)$ and $\lambda^u_{j,s}$ is the summation of time slots which is calculated as 
\begin{align}
\lambda^u_{j,s} =& \ (u - 1)T + \sum_{i=j}^{r} t_i + \sum_{i=1}^{s} t_i. \label{eq:time_slot}
\end{align}
The first term in (\ref{eq:IUI}) is the IUI effect for the transmissions in previous frames. The second term in (\ref{eq:IUI}) is the IUI effect from the transmission in the current frame.

\subsection{Error Probability Analysis} \label{subsec:BER}
In this subsection, we aim to evaluate the error probability in the proposed TDMA-based MCvD system. First of all, we should calculate the distribution of molecules that are counted by node D. Considering the previous subsection, the number of molecules arrived in the volume of node D at the end of the TX-$s$ time slot and frame $n$, denoted by $M^s_{\text{Tot}}[n]$, is 
\begin{align}
M^s_{\text{Tot}}[n] = M^s[n] + M^s_{\text{IUI}}[n]. \label{eq:Total_pre}
\end{align}
By adopting (\ref{eq:current_normal}) and (\ref{eq:IUI}), $M^s_{\text{Tot}}[n]$ is calculated as
\begin{align}
\begin{split}
M^s_{\text{Tot}}[n] \sim & \ \mathcal{N}\bigg(x_s[n]A_s P_h(\mathbf{X_s},t_s) \ , \ x_s[n]A_s P_h(\mathbf{X_s},t_s)  (1 - P_h(\mathbf{X_s},t_s)) \bigg) \\&+ \sum_{u = 1}^{U} \sum_{j=1}^{r} \mathcal{N}\bigg( A_j \ x_j[n - u]  Y^u_{j,s} , A_j \ x_j[n - u]  Y^u_{j,s}  \big( 1 - Y^u_{j,s}\big) \bigg) \\&+ \sum_{j=1}^{s-1} \mathcal{N}\bigg(A_j \ x_j[n] H_{j,s}, A_j \ x_j[n] H_{j,s} \big(1-  H_{j,s} \big)\bigg).
 \label{eq:Total_tot}
\end{split}
\end{align}
Therefore, the distribution of received molecules for TX-$s$ follows the normal distribution given by
\begin{align}
&\text{Pr}(M^s_{\text{Tot}}[n] \mid x_s[n]=0) \sim \mathcal{N}\big( \mu_{0_s},\sigma^2_{0_s} \big), \label{eq:distribution_1}\\
&\text{Pr}(M^s_{\text{Tot}}[n] \mid x_s[n]=1) \sim \mathcal{N}\big( \mu_{1_s},\sigma^2_{1_s} \big). \label{eq:distribution_2}
\end{align}
 The mean and the variance of $M^s_{\text{Tot}}[n]$ are calculated from (\ref{eq:Total_tot}) as follows:
\begin{align}
\begin{split}
\mu_{0_s} =& \ 0.5 \sum_{u = 1}^{U} \sum_{j=1}^{r} A_j Y^u_{j,s} + 0.5 \sum_{j=1}^{s-1} A_j H_{j,s}, \label{eq:mu_0} 
\end{split}
\\
\begin{split}
\mu_{1_s} =& \ A_s P_h(\mathbf{X_s},t_s) + \mu_{0_s}, \label{eq:mu_1}
\end{split}
\\
\begin{split}
\sigma^2_{0_s} =&  \ 0.5 \sum_{u = 1}^{U} \sum_{j=1}^{r} \big\{A_j Y^u_{j,s} - A_j (Y^u_{j,s})^2 \big( 0.5 - 1.25 A_j \big) \big\} \\& + 0.5 \sum_{j=1}^{s-1} \big\{ A_j H_{j,s} - A_j (H_{j,s})^2 \big( 0.5 - 1.25 A_j \big) \big\}, \label{eq:var_0} 
\end{split}
\\
\begin{split}
\sigma^2_{1_s} =& \ A_s P_h(\mathbf{X_s},t_s)  (1 - P_h(\mathbf{X_s},t_s)) + \sigma^2_{0_s}. \label{eq:var_1}
\end{split}
\end{align}
The details of calculating the mean and the variance of $M^s_{\text{IUI}}[n]$ are provided in Appendix~\ref{Appendix I}. 

In this paper, the detection method is based on the Maximum Likelihood (ML) technique. Our system model follows the TDMA technique, thus, the receiver node D decides based on the threshold value for each transmitter node as
\begin{align}
\begin{split}
\hat{x}_{s}[n] = \begin{cases} 
1 & \mbox{if } M^s_{\text{Tot}}[n] \mbox{  $ \ge \tau_s $}, \  \\0 & \mbox{if } M^s_{\text{Tot}}[n]  \mbox{  $< \tau_s $}, \label{eq:rec_r}
\end{cases}
\end{split}
\end{align}
where $\tau_s$ and $\hat{x}_{s}[n]$ are the threshold value at node D for detecting the information sent by the transmitter node TX-$s$, and the information bit detected by the receiver when TX-$s$ is the transmitter, respectively. Note that, we assume that the transmitter nodes and the receiver node are synchronized with each other~\cite{noel2014improving,kilinc2013receiver}. The details of calculating $\tau_s$ is studied in the literature~\cite{kilinc2013receiver}

Now we can calculate the error probability for each TX-$s$. TX-$s$ sends the information at the beginning of its time slot in frame $n$, and the node D detects the information based on (\ref{eq:rec_r}) at the end of the time slot. The error probability when TX-$s$  is the transmitter and node D is the receiver, is given by
\begin{align}
\begin{split}
P^s_{\text{e}}[n]= \ &\text{Pr}(x_s[n] = 1) \cdot \text{Pr}(\hat{x}_s[n]=0\mid x_s[n]=1) +\text{Pr}(x_s[n] = 0) \cdot \text{Pr}(\hat{x}_s[n]=1\mid x_s[n]=0). \label{eq:BER_pre}
\end{split}
\end{align}

We can write the right side of (\ref{eq:BER_pre}) as follows: 
\begin{align}
\begin{split}
\text{Pr}(\hat{x}_s[n]=0\mid x_s[n]=1) =& \ \text{Pr}\big(M^s_{\text{Tot}}[n] < \tau_s \mid x_s[n]=1\big) = \ \frac{1}{2}\bigg( 1 + \text{erf}\bigg( \frac{\tau_s - \mu_{0_s}}{\sqrt{2\sigma_{0_s}^2}} \bigg) \bigg) , \label{eq:rec_0}
\end{split}
\\
\begin{split}
\text{Pr}(\hat{x}_s[n]=1\mid x_s[n]=1) =& \ \text{Pr}\big(M^s_{\text{Tot}}[n] \ge \tau_s \mid x_s[n]=1\big) = \ \ \frac{1}{2}\bigg( 1 - \text{erf}\bigg( \frac{\tau_s - \mu_{1_s}}{\sqrt{2\sigma_{1_s}^2}} \bigg) \bigg). \label{eq:rec_1}
\end{split}
\end{align}

Therefore, the BER for TX-$s$ at the $n^{\text{th}}$ frame is derived as follows~\cite{rudsari2019non}:
\begin{align}
\begin{split}
P^s_{\text{e}}[n]=\frac{1}{2}+\frac{1}{4}\bigg[\text{erf}\bigg(\frac{\tau_s-\mu_{1_s}}{\sqrt{2\sigma_{1_s}^2}}\bigg)-\text{erf}\bigg(\frac{\tau_s-\mu_{0_s}}{\sqrt{2\sigma_{0_s}^2}}\bigg)\bigg]. \label{eq:probability}
\end{split}
\end{align} 

\section{Dynamic TDMA Optimization} \label{sec:TDMA}
In this section, we exploit MOP to minimize the BER of each transmitter TX-$s$ and the receiver D by jointly determining the optimal number of molecules each transmitter node should release and the symbol duration of each time slot meaning that our system is dynamic in terms of the number of allocated molecules and symbol durations. We construct two vectors of variables. The first one is the vector of the number of molecules each transmitter node TX-$s$ releases in its time slot. The second one is the vector of the symbol duration of each time slot. We propose four cases to design the aforementioned system each of which is explained in details in the following subsections.

\subsection{Static-Time Static-Number of Molecules (STSN)} \label{subsec:STSN}
We assume that each transmitter node has equal time and number of molecules for releasing into the medium. We call this approach as STSN and is useful in  MCvD systems that do not require high BER performance. STSN is applicable in systems that cannot tolerate high complexity. The MCvD systems which utilize the pre-encoded transmitter as the transmitter node can be categorized as the systems with low complexity. On the other hand, the MCvD systems which utilize the generic transmitter as the transmitter node can be categorized as the systems that can tolerate higher complexity~\cite{chude2019nanosystems}. To design the STSN-based MC system, we allocate uniform number of molecules and time duration to each transmitter node. By limiting the budget of molecules to $Q$, the allocated number of molecules to each transmitter node is given by
\begin{align}
A_s = \frac{Q}{r}. \label{eq:STSN_A}
\end{align}

The other variable vector in STSN-based MCvD is the symbol duration. Since the system is based on STSN, the time for sending the molecules into the medium is equal for each transmitter node. We assume that the frame is $T$ seconds, i.e., $T = \sum_{s=1}^{r} t_s$. Hence, the symbol duration of each transmitter node is
\begin{align}
t_s = \frac{T}{r}. \label{eq:STSN_t}
\end{align}

\subsection{Dynamic-Time Static-Number of Molecules (DTSN)} \label{subsec:DTSN}
In this approach, we allocate the uniform number of molecules to each transmitter node as (\ref{eq:STSN_A}). The time slot durations should be optimized to minimize the BER for each transmitter node. DTSN can be also applicable MCvD systems with low complexity, due to its simple optimizations solution. We adopt MOP to optimize the vector of symbol durations. The formulated MOP is given by
\begin{align} \label{eq:DTSN_t}
\min_{\boldsymbol{t}}  \ \boldsymbol{P}_e(\boldsymbol{A},\boldsymbol{t}) = [P^1_{e} (\boldsymbol{A},\boldsymbol{t}), ..., P^r_{e} (\boldsymbol{A},\boldsymbol{t})]^T, \   
 \text{s.t.}: \text{C1:}\  t_s > \psi_t,  \ \text{C2:} \ T \le T_{\text{max}},
\end{align}
where C1 and C2 in (\ref{eq:DTSN_t}) are the constraints of the optimization problem. It is worth noting that $s = 1, 2, ..., r$. $\boldsymbol{A}$ and $\boldsymbol{t}$ are the vector of the number of the molecules allocated to each transmitter, i.e., $\boldsymbol{A} = [A_1, A_2, ..., A_r]$ and the vector of symbol durations, i.e., $\boldsymbol{t} = [t_1, t_2, ..., t_r]$, respectively.
 Furthermore, $\psi_t$ is the lower bound of the symbol duration variables. C2 in (\ref{eq:DTSN_t}) means that the total time duration of each time slot should be equal or lower than the maximum frame duration denoted by $T_{\text{max}}$ which is adjusted by the desired application. For example, the time of releasing the drugs plays an important role in drug release mechanism for DDS~\cite{chahibi2013molecular}.

\subsection{Static-Time Dynamic-Number of Molecules (STDN)} \label{subsec:STDN}
In STDN scheme, we assign uniform values for symbol duration as (\ref{eq:STSN_t}). This approach is also of low complexity and provides better BER performance in comparison with STSN and DTSN. The performance comparison of introduced approaches has been elevated in Section~\ref{sec:numerical_analysis}. The resource allocation problem is defined via an MOP as follows:
\begin{align}\label{eq:STDN_A}
\min_{\boldsymbol{A}}   \ \boldsymbol{P}_e(\boldsymbol{A},\boldsymbol{t}) = [P^1_{e} (\boldsymbol{A},\boldsymbol{t}), ..., P^r_{e} (\boldsymbol{A},\boldsymbol{t})]^T, \ \ \   
 \text{s.t.}: \text{C1:} \ \psi_A < A_s < \Psi_A .
\end{align} 
$\psi_A $ and $\Psi_A $ are the lower and the upper bounds of the number on molecules allocated to each transmitter, respectively. C1 in (\ref{eq:STDN_A}) is enforced by the requirements of the application. For example, minimizing the drug dosage in Human Immunodeficiency Virus (HIV) therapy by molecular communication in novel DDS has a crucial importance~\cite{chahibi2013molecular}.

\subsection{Dynamic-Time Dynamic-Number of Molecules (DTDN)} \label{subsec:DTDN}
As the last approach, we optimize both of the symbol durations and the number of molecules allocated to each transmitter. Despite the BER of the system is better than other approaches, DTDN is applicable in MCvD systems with high complexity. In addition, this approach is applicable in release management mechanism for novel DDSs that controls simultaneously the drug dosage and their releasing time. The optimization problem to minimize the BER values of each transmitter is given by
\begin{subequations}\label{eq:DTDN}
\begin{align}
\begin{split}\label{eq:DTDN_1}
\min_{\boldsymbol{A},\boldsymbol{t}}  \ &\boldsymbol{P}_e(\boldsymbol{A},\boldsymbol{t}) = [P^1_{e} (\boldsymbol{A},\boldsymbol{t}), ..., P^r_{e} (\boldsymbol{A},\boldsymbol{t}) ]^T,   
\end{split}
\\
\begin{split}\label{eq:DTDN_2}
 \text{s.t.}: \ &\text{C1:} \ \psi_A  < A_s < \Psi_A , \ \ \text{C2:}  \ t_s > \psi_t, \ \ \text{C3:} \ T \le T_{\text{max}}, 
\end{split}
\end{align}
\end{subequations}
where C1 and C2 are the constraints that define the lower and the upper bounds of the number of molecules and the lower bound of symbol durations allocated to each transmitter node, respectively. C3 in~\ref{eq:DTDN_2} defines the maximum frame time constraint. 
To solve the optimization problem introduced in (\ref{eq:DTSN_t}), (\ref{eq:STDN_A}), and (\ref{eq:DTDN}), we employ MOP whose details are given in the next section.

\section{Multi-objective Optimization Solution} \label{sec:multi_objective_optimization}
The solution of MOP is considered as the values which are in the Pareto frontier set~\cite{messac2003normalized}. To solve MOP, the authors have proposed some mathematical techniques which we exploit two categories of them. The first one is the Pareto method, which keeps the vector of variables independent during the optimization. If we consider two objective functions as MOP, the optimal values in MOP can be obtained when one objective function has no increase without reducing the other objective function. This case is called the Pareto optimal or non-dominated solution. The other one is called non-Pareto optimal solution which is based on the scalarization method (SM). The set of optimal solutions, in this case, is referred to as Pareto optimal solutions~\cite{ehrgott2005multicriteria}. The goal of SM is taking MOP into scalar fitness function as~\cite{murata1996multi}
\begin{align}
G(\boldsymbol{x}) = \sum_{u=1}^{r}w_u g_u(\boldsymbol{x}), \label{eq:WSM}
\end{align}
where $g_v$ and $w_v$ are the objective function and the weight of it, respectively. There are some methods to assign weights to objective functions, such as equal weights, Rank Order Centroid (ROC) weights, and Rank-sum (RS) weights~\cite{dawes1974linear,einhorn1977simple}.

In this paper, we adopt WSM to solve the multi-objective optimization problem, because its complexity is lower than other solutions~\cite{marler2010weighted,kim2006adaptive,madsen2009groupwise}. If all weights in (\ref{eq:WSM}) be positive, by minimizing $G(\boldsymbol{x})$, a sufficient condition for Pareto optimally is reached, i.e, the minimum of (\ref{eq:WSM}) is always Pareto optimal~\cite{goicoechea1982multiobjective,zadeh1963optimality}. We use equal weights and by considering $\sum_{u=1}^{r} w_u = 1$ where the weights become~\cite{marler2010weighted,dawes1974linear}
 \begin{align}
 w_u = \frac{1}{r}.
 \end{align}
 Therefore, (\ref{eq:DTSN_t}), (\ref{eq:STDN_A}), and (\ref{eq:DTDN}) become single objective optimization problems with equal weights. 

 \subsection{DTSN Optimization Solution} \label{subsec:DTSN_solution} 
 The optimization problem to find the optimal values of $\boldsymbol{t}^*$, which is introduced in (\ref{eq:DTSN_t}), yields the minimization of BER subject to constraints regarding the lower bound of $t_s$ and the maximum frame time duration of $T_{\text{max}}$ as given by
 \begin{align}\label{eq:DTSN_t_single}
\min_{\boldsymbol{t}} \   \boldsymbol{P}_e(\boldsymbol{A},\boldsymbol{t}) = \frac{1}{r} \sum_{s=1}^{r} P^s_{e} (\boldsymbol{A},\boldsymbol{t}),  \ \ \ 
 \text{s.t.}: \ \text{C1:} \  t_s > \psi_t,  \ \text{C2:} \ T \le T_{\text{max}}.
 \end{align}
The objective function in (\ref{eq:DTSN_t_single}) is convex on each time slot duration for TX-$s$, i.e., $t_s$, as proved in Appendix~\ref{Appendix III}. To solve (\ref{eq:DTSN_t_single}), we utilize Alternative Search Method (ASM)~\cite{moltafet2018joint}. In this regard, the DTSN optimization problem is divided into $r$ sub-problems and each of them are solved by the publicly available software CVX~\cite{grant2015cvx} to converge to a sub-optimal solution. The results of the algorithm to solve (\ref{eq:DTSN_t_single}) based on ASM is 
\begin{align}
\begin{split}  \Rightarrow
\{t_1[0] \rightarrow ... \rightarrow t_r[0] \} 
\Rightarrow ... \Rightarrow \{ t_1[\alpha_{\text{opt}-1}] \rightarrow ... \rightarrow t_r[\alpha_{\text{opt}-1}]  \} 
\Rightarrow \{ t_1[\alpha_{\text{opt}}] \rightarrow ... \rightarrow t_r[\alpha_{\text{opt}}]  \}, \label{eq:DTSN_ASM}
\end{split}
\end{align}
where $\{t_1[0] \rightarrow ... \rightarrow t_r[0] \}$ is the initial setting values for the optimization problem (\ref{eq:DTSN_t_single}). In the beginning of each iteration $\alpha$, $t_1[\alpha]$ is obtained given the previous values of other time slot durations, and the same argument for obtaining other time slot durations is applicable. The algorithm is iterated until some convergence criteria is satisfied. Note that $\alpha_{\text{opt}}$ is the number of iterations needed to converge to the sub-optimal solution.

\subsection{STDN Optimization Solution} \label{subsec:STDN_solution}
The optimization problem to find the optimum number of molecules denoted by $\boldsymbol{A}^*$ allocated to each transmitter in STDN introduced in (\ref{eq:STDN_A}) becomes a single objective optimization problem as given by
\begin{align}  \label{eq:STDN_A_single}
\min_{\boldsymbol{A}} \  &\boldsymbol{P}_e(\boldsymbol{A},\boldsymbol{t}) = \frac{1}{r} \sum_{s=1}^{r} P^s_{e} (\boldsymbol{A},\boldsymbol{t}), \   
 \text{s.t.}: \ \text{C1:} \  \psi_A  < A_s < \Psi_A . 
\end{align}
The objective function in (\ref{eq:STDN_A_single}) is convex on $A_s$, and therefore, the optimization problem (\ref{eq:STDN_A_single}) is convex on $A_s$. The proof of convexity the optimization problem (\ref{eq:STDN_A_single}) is provided in Appendix~\ref{Appendix IV}. It is worth noting that we consider $\boldsymbol{A}$ as a real variable and after optimizing it, we quantize $\boldsymbol{A}$ to the nearest integer value. The discussion of BER difference between considering $\boldsymbol{A}$ as real and integer variable is provided in Section~\ref{sec:numerical_analysis}.

The optimization problem (\ref{eq:STDN_A_single}) is solved by utilizing ASM in which is divided into $r$ sub-problems. Each sub-problem is solved by the publicly available software CVX. The ASM algorithm iteratively solves each sub-problem to converge to a sub-optimal solution. The ASM algorithm is given by
\begin{align}
	\begin{split}  \Rightarrow
		\{A_1[0] \rightarrow ... \rightarrow A_r[0] \} 
		\Rightarrow ... \Rightarrow \{ A_1[\beta_{\text{opt}-1}] \rightarrow ... \rightarrow A_r[\beta_{\text{opt}-1}]  \} 
		\Rightarrow \{ A_1[\beta_{\text{opt}}] \rightarrow ... \rightarrow A_r[\beta_{\text{opt}}]  \}, \label{eq:STDN_ASM}
	\end{split}
\end{align}
where $\{A_1[0] \rightarrow ... \rightarrow A_r[0] \}$ and $\beta_{\text{opt}}$ are the initial setting values for the STDN optimization problem and the number of iterations needed to converge to the sub-optimal solution, respectively.

\subsection{DTDN Optimization Solution} \label{subsec:DTDN_solution}
The DTDN approach is also a MOP which can be solved by taking the WSM into account. Problem (\ref{eq:DTDN}) becomes a single objective optimization problem that calculates the optimum symbol duration denoted by $\boldsymbol{t}^*$ and the number of allocated molecules to each transmitter $\boldsymbol{A}^*$, simultaneously, which is given as follows:
\begin{subequations}\label{eq:DTDN_single}
\begin{align}
\begin{split}\label{eq:DTDN_single_1}
\min_{\boldsymbol{A},\boldsymbol{t}}  \  &\boldsymbol{P}_e(\boldsymbol{A},\boldsymbol{t}) = \frac{1}{r} \sum_{s=1}^{r} P^s_{e} (\boldsymbol{A},\boldsymbol{t})  
\end{split}
\\
\begin{split}\label{eq:DTDN_single_2}
\text{s.t.}: \ &\text{C1:} \ \psi_A  < A_s < \Psi_A ,  \ \ \ 
\text{C2:} \  t_s > \psi_t , \ \ \ 
\ \text{C3:} \ T \le T_{\text{max}}. 
\end{split}
\end{align}
\end{subequations}
To solve (\ref{eq:DTDN_single}), we employ ASM. By utilizing it, (\ref{eq:DTDN_single}) is divided into two sub-problems as follows:
\begin{subequations}\label{eq:DTDN_sub_A}
	\begin{align}
	\begin{split}\label{eq:DTDN_sub_A_1}
		\min_{\boldsymbol{A}} \   &\boldsymbol{P}_e(\boldsymbol{A},\boldsymbol{t}) = \frac{1}{r} \sum_{s=1}^{r} P^s_{e} (\boldsymbol{A},\boldsymbol{t}),  \ \ \
		\ \text{s.t.}: \ \text{C1:} \ \psi_A < A_s < \Psi_A,
	\end{split}
	\\
	\begin{split}\label{eq:DTDN_sub_t}
		\min_{\boldsymbol{t}}  \  &\boldsymbol{P}_e(\boldsymbol{A},\boldsymbol{t}) = \frac{1}{r} \sum_{s=1}^{r} P^s_{e} (\boldsymbol{A},\boldsymbol{t}),  \ \ \ \
		 \text{s.t.}: \  \text{C1:} \  t_s > \psi_t, \ \text{C2:} \ T \le T_{\text{max}}. 
	\end{split}
	\end{align}
\end{subequations}
Based on ASM, the optimization problems (\ref{eq:DTDN_sub_A_1}) and (\ref{eq:DTDN_sub_t}) are iteratively solved until it converges to a sub-optimal solution. This technique is provided by the following algorithm
\begin{align}
	\begin{split}  \Rightarrow
	\{\boldsymbol{A}[0] \rightarrow \boldsymbol{t}[0]\} 
	\Rightarrow ... \Rightarrow \{ \boldsymbol{A}[\gamma_{\text{opt}-1}] \rightarrow \boldsymbol{t}[\gamma_{\text{opt}-1}]  \} 
	 \Rightarrow \{ \boldsymbol{A}[\gamma_{\text{opt}}] \rightarrow \boldsymbol{t}[\gamma_{\text{opt}}]  \}, \label{eq:ASM}
	\end{split}
\end{align}
where $\{\boldsymbol{A}[0] ,\boldsymbol{t}[0] \}$ is the initial setting values for the optimization problem (\ref{eq:DTDN_single}) and $\gamma_{\text{opt}}$ is the number of required iterations to converge to the sub-optimal solution.

\section{Computational Complexity} \label{sec:comp_complexity}
In this section, the computational complexity of the optimization problems is provided. 
The alternate method is employed in which two sub-problems are solved in each iteration in DTDN approach as~I)~finding the symbol durations by utilizing ASM for $r$ sub-problems and~II)~allocation of the number of molecules to each transmitter by utilizing ASM for $r$ sub-problems. The complexity of ASM is linearly proportional to the number of iterations needed to converge and the complexity of the solution of each sub-problem~\cite{moltafet2018joint}. The CVX toolbox is used to solve each sub-problems of optimization problems~(\ref{eq:DTSN_t_single}) and (\ref{eq:STDN_A_single}). The CVX toolbox exploits the interior point method to solve the convex optimization problem. The computational complexity of the CVX toolbox is given by $\mathcal{C} = \dfrac{\log (\frac{\Lambda}{\rho_1 \rho_2})}{\log \rho_3}$~\cite{grant2015cvx,mokari2016limited}, where $\Lambda$ is the number of constraints in each problem, $\rho_1$ is the initial point to approximate the accuracy of the interior point method, $\rho_2$ is the stopping criterion of the interior point method, and $\rho_3$ is used to update the accuracy of the interior point method~\cite{grant2015cvx}. 
All the computational complexities of the proposed approaches are provided in Table~\ref{Table:complexity}. 

\begin{table}[t]
	\centering
	\caption{Computational Complexity of The Proposed Approaches}
	\begin{tabular}{c c}
		Approach & Computational Complexity\\ \hline
		DTSN &$ r  \alpha_{\text{opt}}\dfrac{\log (\frac{\Lambda}{\rho_1 \rho_2})}{\log \rho_3}$ \\ \hline
		STDN & $ r  \beta_{\text{opt}}\dfrac{\log (\frac{\Lambda}{\rho_1 \rho_2})}{\log \rho_3}$ \\ \hline
		DTDN & $2 r^2 \alpha_{\text{opt}}  \beta_{\text{opt}}  \gamma_{\text{opt}} \dfrac{\log (\frac{\Lambda}{\rho_1 \rho_2})}{\log \rho_3}$\\ \hline
	\end{tabular}
	\label{Table:complexity}
\end{table}

  \begin{table}[]
	\caption{Values and Ranges of TDMA-based MCvD System Parameters }
	\centering
	\tiny
	\begin{tabular}{|l|l|l|}
		\hline
		Parameter& Variable & Values \\ \hline
		\hline
		Diffusion Coefficient&$\Omega$ &$\{\text{3.7}, \text{4.5}, \text{4.87}\} \  \times \text{10}^{\text{-9}} \text{m}^\text{2}$/s~\cite{cussler2009diffusion}\\
		\hline
		Drift Velocity&$(u_x, u_y, u_z)$                        &     $(\text{100}, \text{200}, \text{100}) \mu$m/s~\cite{bhatnagar20193}\\
		\hline
		The number of transmitters &  $r$ &    3\\ 
		\hline
		Location of TX-1 &  $(X_{11}, X_{21}, X_{31})$ &  $(\text{65}, \text{20}, \text{30}) \mu$m\\ 
		\hline
		Location of TX-2 &  $(X_{12}, X_{22}, X_{32})$  &  $(\text{60}, \text{10}, \text{30}) \mu$m\\ 
		\hline
		Location of TX-3 &  $(X_{13}, X_{23}, X_{33})$  &    $(\text{50}, \text{10}, \text{30}) \mu$m\\ 
		\hline
		Location of node D &  $(D_x, D_y, D_z)$ &    $(\text{100}, \text{20}, \text{40}) \mu$m\\ 
		\hline
		Lower and upper bounds of number of molecules &  $\psi_A, \Psi_A$ &    $\{\text{100}, \text{800}\}$\\ 
		\hline
		Lower bound of symbol duration &  $\psi_t$ &     $\text{1}\mu\text{s}$\\ 
		\hline
		The IUI length	           &   $U$                              &     3\\
		\hline
		The radius of the receiver	           &   $d$                              &     45$\mu$m~\cite{bhatnagar20193}\\
		\hline
	\end{tabular}
	\label{table}
\end{table}

\section{Numerical Analysis} \label{sec:numerical_analysis}
In this section, we present the numerical analysis of the dynamic TDMA-based MCvD system. In order to evaluate the error probability, we apply four approaches mixed by the static and dynamic behavior for symbol duration and the number of molecules. Furthermore, the aforementioned approaches are analyzed in case of considering the MDE, MODE, and SDE scenarios. We consider 3 transmitter nodes which are placed at different distances from the destination node D. The locations of the transmitter nodes are provided in Table \ref{table}. We consider 3 scenarios for the environment and information molecules which are discussed in Section~\ref{sec:system_model}.

	\begin{figure*}%
	\centering
	\subfigure[]{%
		\label{Fig:Mixture_mild}%
		\includegraphics[height=120pt]{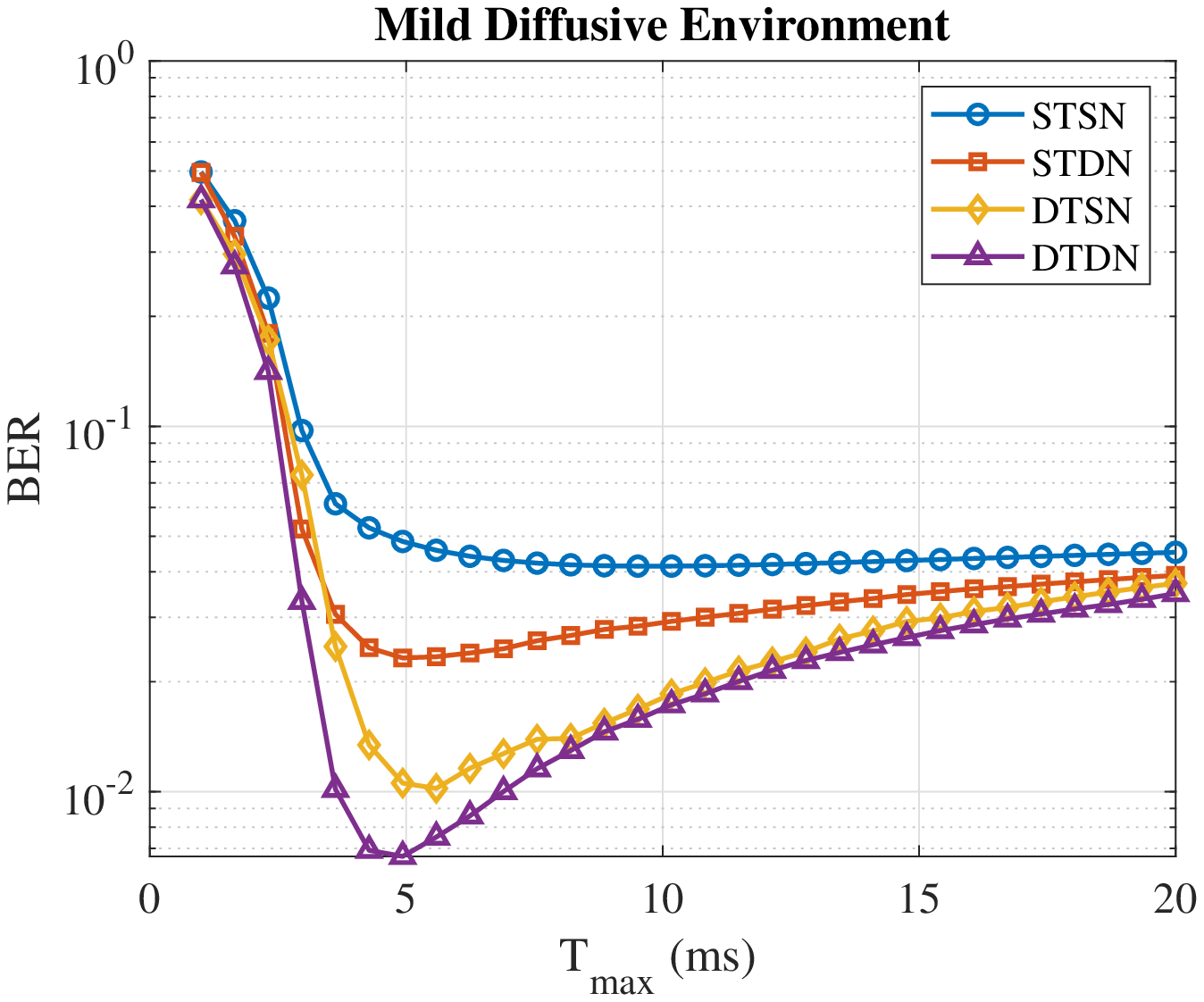}}%
	\subfigure[]{%
		\label{Fig:Mixture_moderate}%
		\includegraphics[height=120pt]{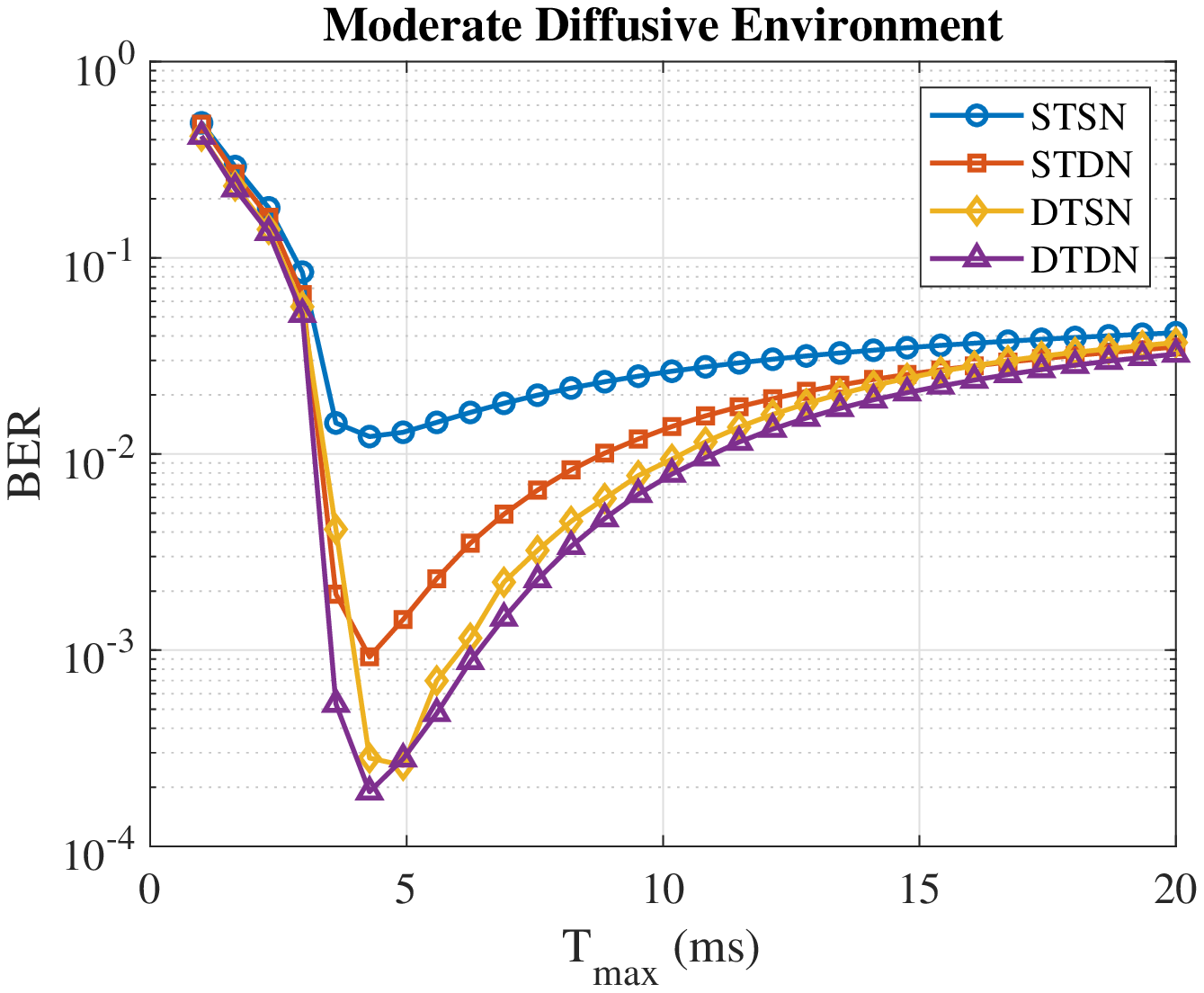}}%
	\subfigure[]{%
		\label{Fig:Mixture_severe}%
		\includegraphics[height=120pt]{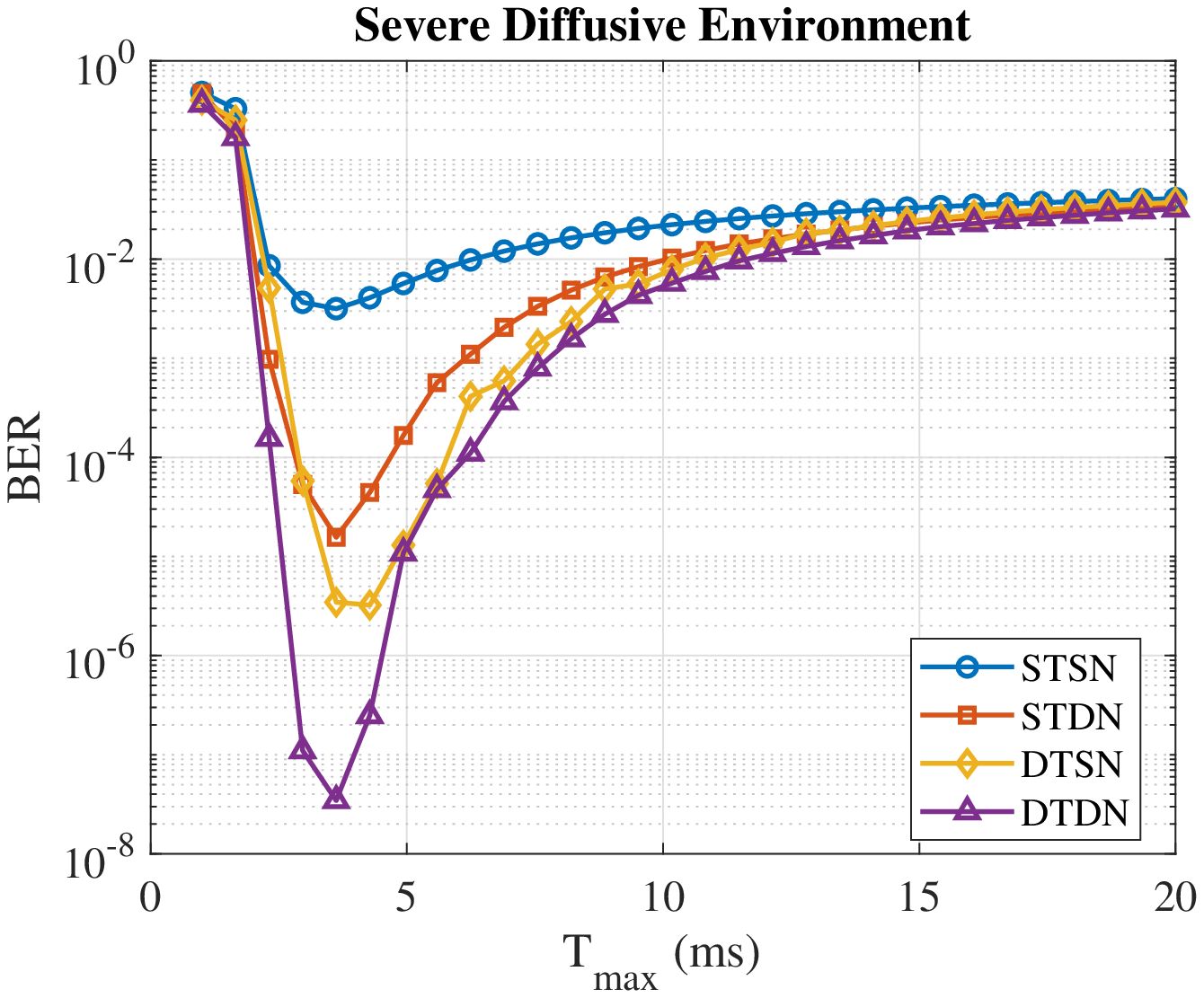}}%
	\caption{ The minimized BER of the TDMA-based MCvD system as a function of frame for different approaches and scenarios as: (a) MDE, (b) MODE, and (c) SDE.  }
	\label{fig:4}
\end{figure*}

The first scenario is based on MDE where the Carbon tetrachloride molecules inside the n-Heptane medium is considered. In this scenario, the probability of reception the molecules in the destination node is lower than other scenarios. Therefore, the performance of the system is not high compared to the other scenarios. Fig.~\ref{Fig:Mixture_mild} shows the minimized BER by employing MDE scenario when STSN, STDN, DTSN, and DTDN approaches are considered in dynamic TDMA-based MCvD as a function of $T_{\text{max}}$. It is worth noting that we consider $T=T_{\text{max}}$ in STSN approach.
In performance point of view, DTDN approach is the best among others which is indicated in Fig.~\ref{Fig:Mixture_mild}. 
In addition, the BER is minimized in case of constraining $T_{\text{max}}$ and it is shown that for some particular $T_{\text{max}}$, the minimized BER reaches the maximum performance. For $T_{\text{max}}$~=~4.931~ms, the minimized BER reaches to $\text{6.6}\times\text{10}^\text{-3}$ in DTDN approach. However, by exploiting the STSN approach, the minimum of minimized BER is $\text{4.1}\times\text{10}^\text{-2}$ for $T_{\text{max}}=$~9.517~ms. Another key fact to remember is that STDN approach is better than DTSN for $T_{\text{max}}<\text{3.293~ms}$ from the performance point of view and for $T_{\text{max}}>\text{3.293~ms}$ DTSN approach is always better than STDN. It means that for high values of frame, optimizing symbol durations is better than optimizing the number of molecules allocated to each transmitter node. Therefore, for small values of $T_{\text{max}}$, optimizing the number of molecules is more effective. This result could be applicable in employing the drug release mechanism in novel DDSs. However, the DTDN approach has the best performance but by referring to Table~\ref{Table:complexity}, its complexity is high. On the other hand, the DTSN approach is less complex than DTDN. Hence, DTDN approach is more applicable in MDE with high computational complexity potency (CCP) nanomachines, i.e., the nanomachines that can tolerate high complexity and require BER lower than $\text{10}^{-2}$. In the MDE scenario with moderate CCP nanomachines, we suggest DTSN as the approach of a drug release mechanism in novel DDSs. However, in low CCP nanomachines, we suggest STSN approach in which manages the drug releasing with the lowest complexity, but the error probability is more than other approaches.

In Fig.~\ref{Fig:Mixture_moderate}, we present the performance of the system in terms of minimizing BER as a function of $T_{\text{max}}$ in which STSN, STDN, DTSN, and DTDN approaches are assessed in MODE. In this scenario, we consider the Hydrogen molecules as the information molecules and the water as the medium. The approaches' BER in MODE is similar to MDE but there are some differences. In MODE situation, the probability of reception the molecules is more than that of MDE one, and therefore, BER performance in MODE is better than MDE. This behavior is shown in all the approaches. Moreover, DTDN approach in all $T_{\text{max}}$ values is better than other approaches, because this procedure optimizes the number of molecules and symbol durations simultaneously. However, from computational complexity point of view, by referring to Table~\ref{Table:complexity}, Fig.~\ref{Fig:Mixture_moderate} shows that DTSN approach is better than other approaches, because DTSNs BER is very close to DTDN, but, in spite of that, the DTSNs complexity is lower than DTDN. Therefore, in MODE case, we suggest DTSN as the best approach to manage the drug releasing in novel DDSs. However, for $T_{\text{max}}<$~3.948~ms, STDN BER performance is better than DTSN. But for $T_{\text{max}}>$~3.948~ms, DTSN is better than STDN. This behavior in MODE is similar to MDE. The maximum performance of DTDN is attained in $T_{\text{max}}$~=~4.276~ms in which BER is $\text{1.8}\times\text{10}^\text{-4}$, and for STSN approach, the maximum performance is $\text{1.2}\times\text{10}^\text{-2}$ which is better in comparison with MDE scenario. In addition, in this scenario, we suggest STSN as the approach of managing the drug releasing in low CCP nanomachines.

Fig.~\ref{Fig:Mixture_severe} shows the BER performance of the proposed system in SDE. In this scenario, we consider the Propane as information molecules and the n-Heptane as the medium. The probability of receiving information molecules in SDE is better than MDE and MODE. Therefore, the BER performance is better than other scenarios. In SDE scenario, DTDN approach has the best performance for all values of $T_{\text{max}}$. Moreover, all STSN, STDN, DTSN, and DTDN approaches have better performance in comparison utilizing MDE and MODE scenarios. For example, the maximum BER performance in DTDN is $\text{3.45}\times\text{10}^\text{-8}$ which is better than that of MDE and MODE. There is also another important point in which is observed in Fig.~\ref{Fig:Mixture_severe}. The BER in DTDN and DTSN for $T_{\text{max}}>$~5~ms are close together which means for the high value of $T_{\text{max}}>$, we can employ DTSN approach to manage the drug releasing, due to the fact that DTSNs complexity is lower than DTDN (see Table~\ref{Table:complexity}). Furthermore, for $T_{\text{max}}<$~2.966~ms, BER in STDN is better than DTSN, and for $T_{\text{max}}>$~2.966~ms BER in DTSN is better than STDN. It means that for novel DDS with high CCP nanomachines, DTDN approach is suggested to manage the drug releasing mechanism. 
In case of SDE in moderate CCP, for $T_{\text{max}}<$~2.966~ms, STDN is suggested and for $T_{\text{max}}>$~2.966~ms, DTSN is suggested. In all scenarios, for low CCP nanomachines, we suggest STSN approach. However, the error probability in this approach is more than other approaches, but from the complexity point of view, it is the lowest one. We summarize the drug release management in MDE, MODE, and SDE scenarios with different suggested approaches in Table~\ref{Table:DDS}. 

  \begin{table}[]
	\caption{Drug Releasing Management Approaches}
	\centering
	\tiny
	\begin{tabular}{|c|c|c|c|}
		\hline
		Scenario& CCP &Maximum Time Frame& Suggested Approach \\ \hline
		\hline
		\multirow{6}*{MDE} 
		& \multirow{2}*{Low} & \multirow{2}*{$T_{\text{max}}$~=~[1,~20] ms}&\multirow{2}*{STSN} 
		\\ 
		&  &  &  
		\\ \cline{2-4}
		& \multirow{2}*{Moderate} & 1~ms~$<T_{\text{max}}<$~3.293~ms&STDN 
		\\ \cline{3-4}
		&  & 3.293~ms$<T_{\text{max}}<$~20~ms&DTSN 
		\\ \cline{2-4}
		&\multirow{2}*{High}  & \multirow{2}*{$T_{\text{max}}$~=~[1,~20] ms}& \multirow{2}*{DTDN}
		\\ 
		&  &  & 
		\\ \hline
		\multirow{6}*{MODE}
		& \multirow{2}*{Low}  &  \multirow{2}*{$T_{\text{max}}$~=~[1,~20] ms}& \multirow{2}*{STSN} 
		\\ 
		&   &  & 
		\\ \cline{2-4}
		& \multirow{2}*{Moderate} & 3.948~ms~$<T_{\text{max}}<$~14.76~ms&DTSN 
		\\ \cline{3-4}
		&  & 1~ms~$<T_{\text{max}}<$~3.948~ms or 14.76~ms~$<T_{\text{max}}<$~20~ms&STDN 
		\\ \cline{2-4}
		&\multirow{2}*{High}  &  \multirow{2}*{$T_{\text{max}}$~=~[1,~20] ms}& \multirow{2}*{DTDN}
		\\ 
		&  &  & 
		\\ \hline
		\multirow{6}*{SDE} 
		& \multirow{2}*{Low} &  \multirow{2}*{$T_{\text{max}}$~=~[1,~20] ms}& \multirow{2}*{STSN}
		\\ 
		&  &  & 
		\\ \cline{2-4}
		& \multirow{2}*{Moderate} & 2.966~ms~$<T_{\text{max}}<$~13.45~ms&DTSN 
		\\ \cline{3-4}
		&  & 1~ms~$<T_{\text{max}}<$~2.966~ms or 13.45~ms~$<T_{\text{max}}<$~20~ms&STDN
		\\ \cline{2-4}
		&\multirow{2}*{High}  &  \multirow{2}*{$T_{\text{max}}$~=~[1,~20] ms}& \multirow{2}*{DTDN}
		\\
		&  & & 
		\\ \hline
	\end{tabular}
	\label{Table:DDS}
\end{table}

	\begin{figure*}%
	\centering
	\subfigure[]{%
		\label{Fig:STSN_budget}%
		\includegraphics[height=150pt]{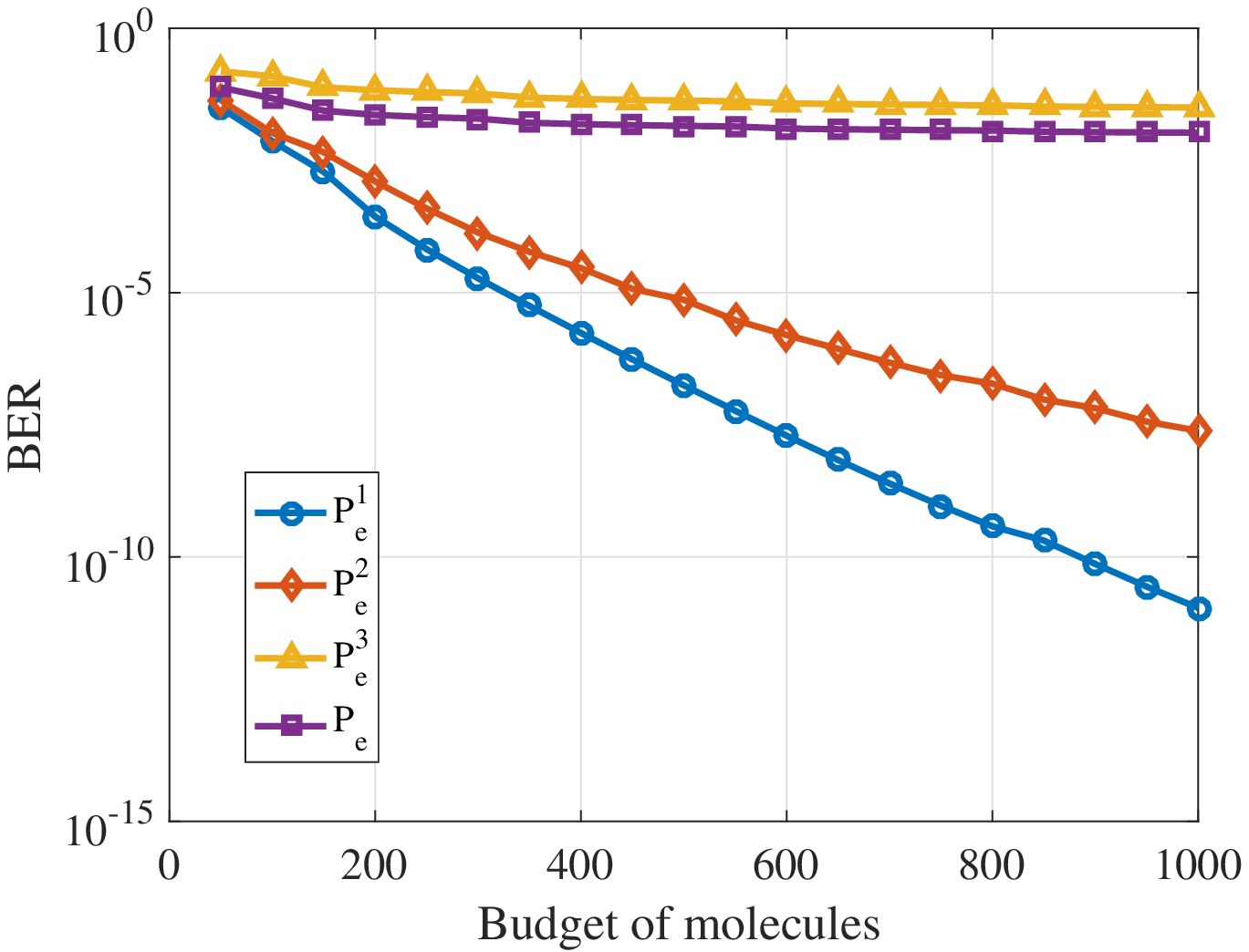}}%
	\subfigure[]{%
		\label{Fig:DTSN_budget}%
		\includegraphics[height=150pt]{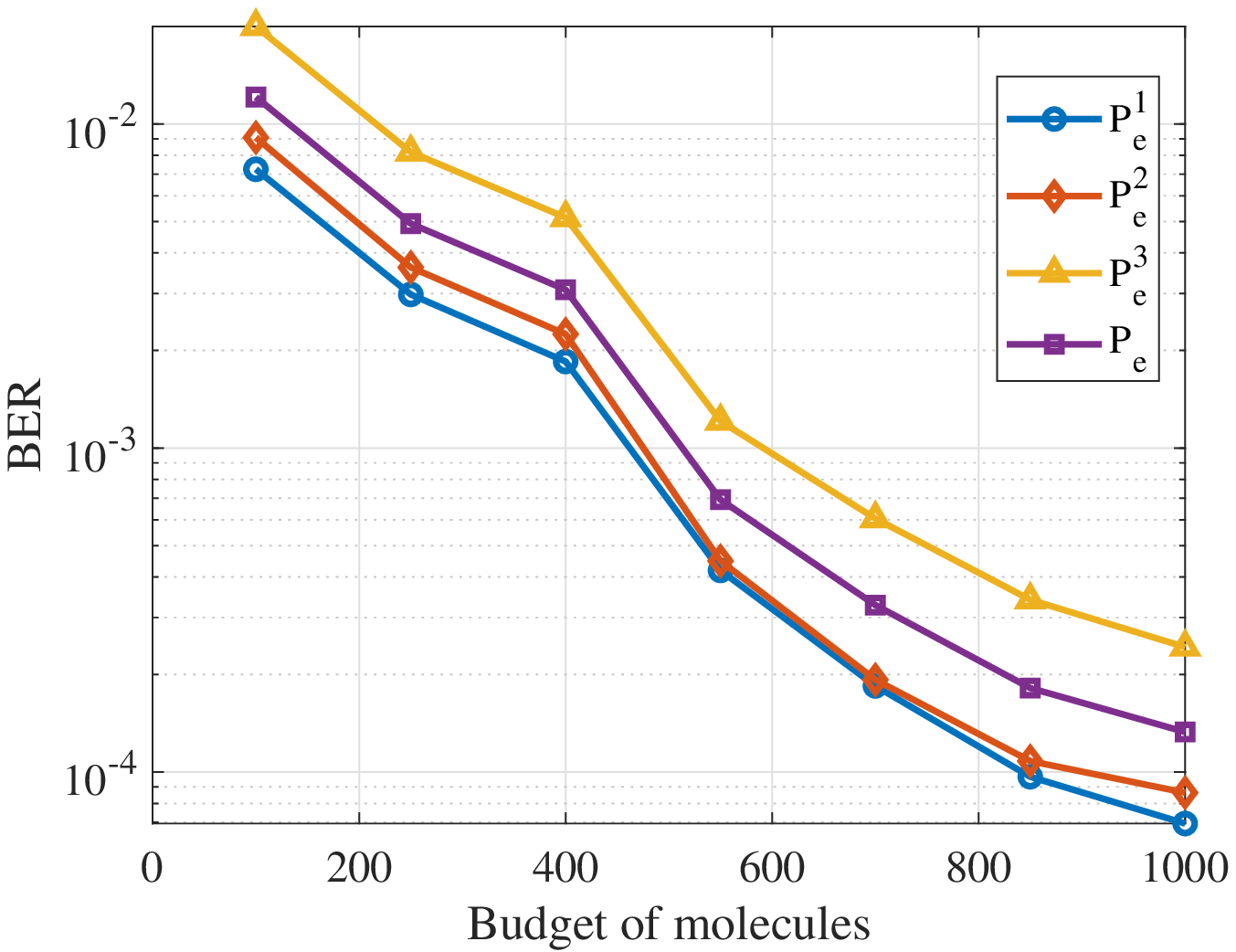}}%
	\caption{The minimized BER of TDMA-based MCvD system as a function of budget of molecules for two approaches as: (a) STSN and (b) DTSN ($T_{\text{max}}$~=~4.95~ms and $\Omega = \text{4.5}\times\text{10}^{\text{-9}}~\text{m}^2/\text{s}$).  }
\end{figure*}

Up to here, we discuss the MDE, MODE, and SDE scenarios to assess the drug releasing management in novel DDS. In the following, we consider MODE scenario as an instance to evaluate how  optimization problems optimize the number of molecules and symbol durations. In Fig.~\ref{Fig:STSN_budget}, we plot the BER performance of STSN approach as a function of different budget of molecules. In this content, the BER of the system which is calculated by WSM is decreased by increasing the budget of molecules. However, BER of TX-1 is better than others due to the fact that TX-1 is closer to the destination node. At the following, we discuss the performance of the system by increasing the budget of molecules in case of utilizing the DTSN approach.


In Fig.~\ref{Fig:DTSN_budget}, we present the optimization performance of BER as a function of the budget of molecules allocated to the transmitters in case of considering DTSN approach. It shows that by increasing the budget of molecules, the BER decreases. It is due to the fact that increasing the budget, increases the number of molecules allocated to each transmitter node by referring to~(\ref{eq:STSN_A}). Moreover, there is a remarkable difference between the performance achieved with STSN approach (see $P_e$ in Fig.~\ref{Fig:STSN_budget}) and DTSN approach by increasing the budget of molecules. As a fair comparison, we set a fixed value of $T_{\text{max}}$ in the analysis of BER for STSN and DTSN. Fig.~\ref{Fig:DTSN_budget} demonstrates that BER for TX-1 is better than other transmitters for all the budgets of molecules because the transmitter node TX-1 is closer to the destination node. The DTSN approach reach the BER of $\text{1.3}\times\text{10}^\text{-4}$. On the other hand, by employing STSN approach, the BER reaches to $\text{10}^\text{-2}$ in which is very lower than the BER provided by DTSN.

	\begin{figure*}%
	\centering
	\subfigure[]{%
		\label{Fig:STDN_bar}%
		\includegraphics[height=150pt]{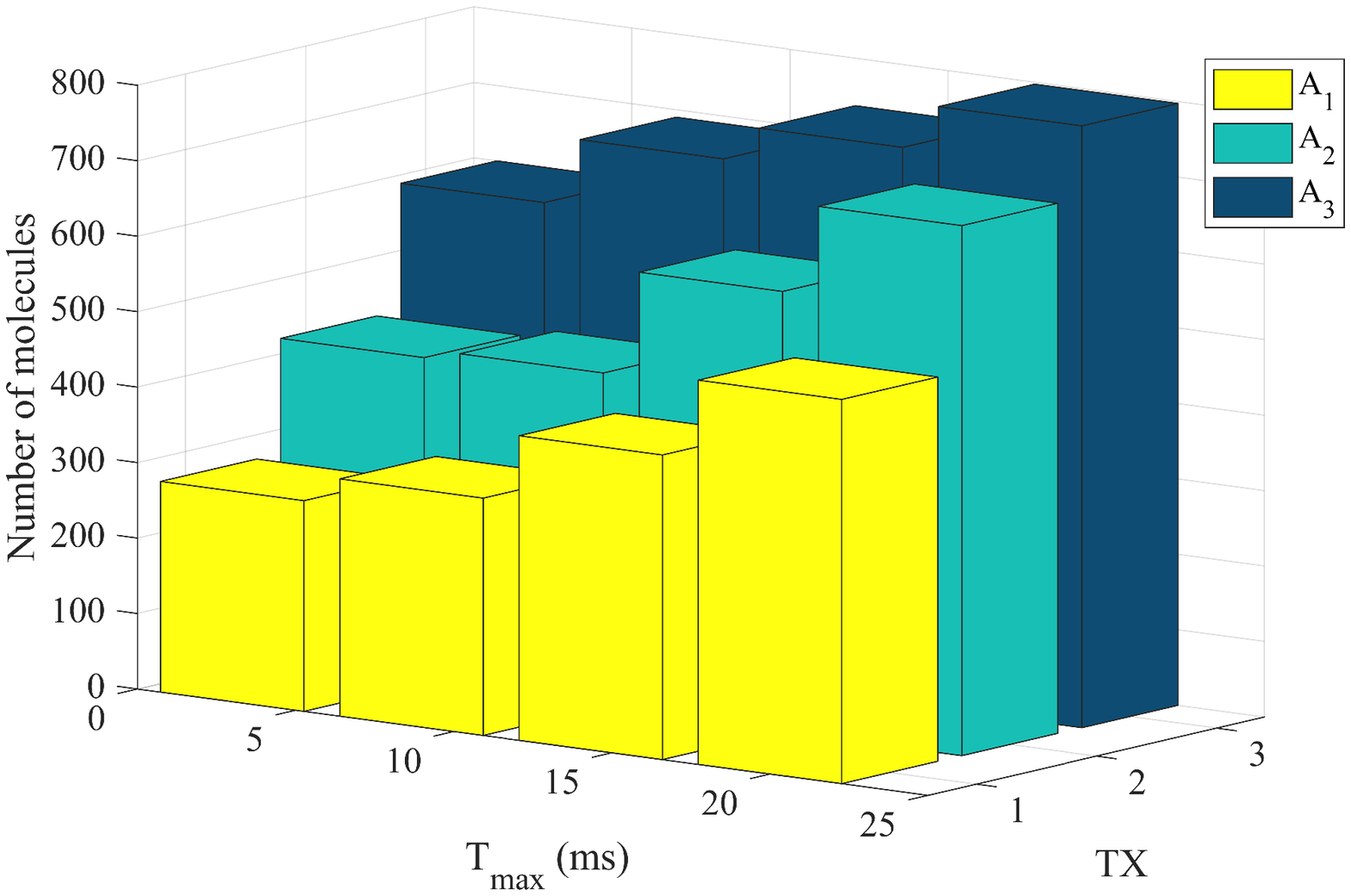}}%
	\subfigure[]{%
		\label{Fig:DTSN_frame_bar}%
		\includegraphics[height=150pt]{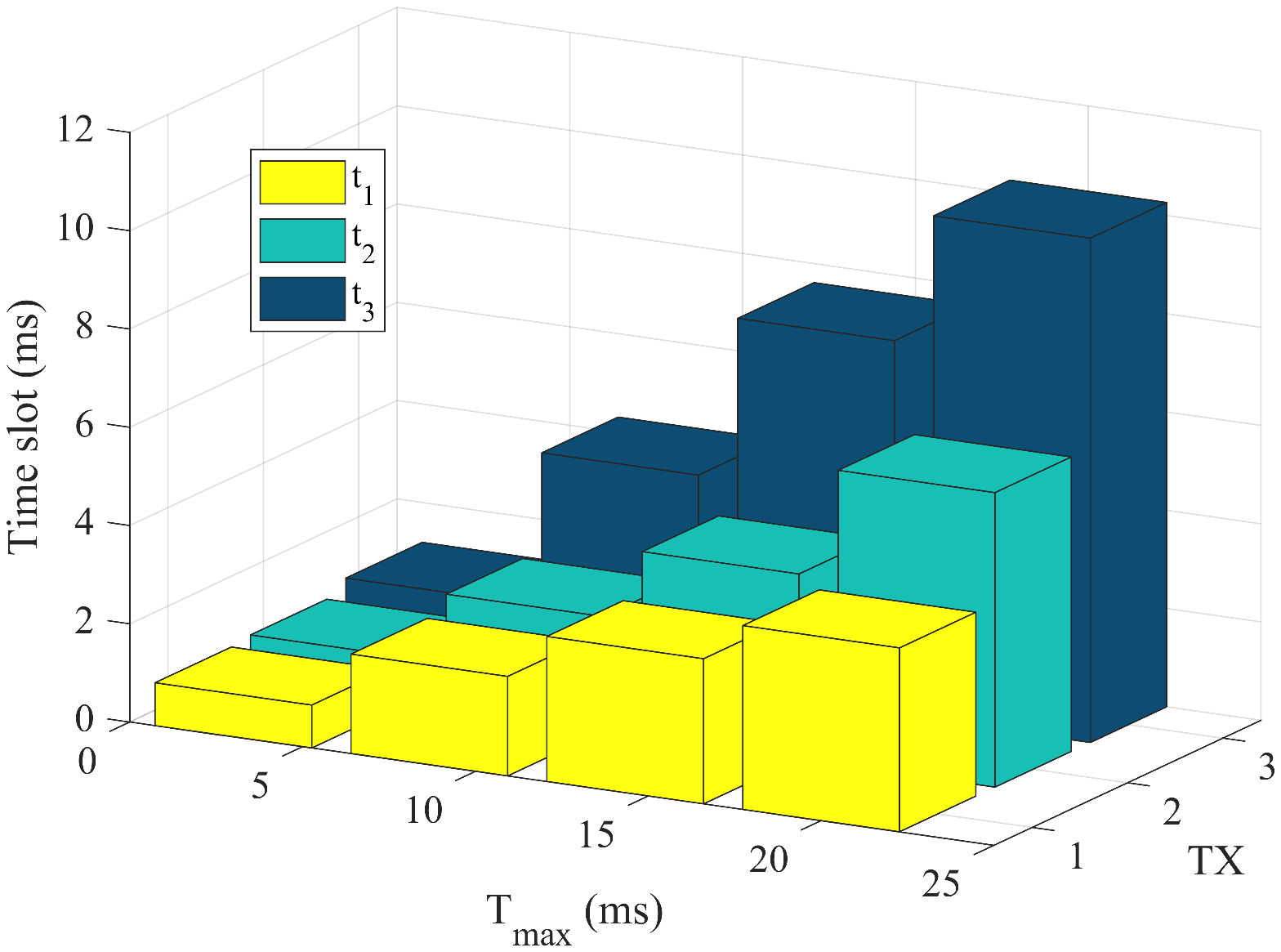}}%
	\caption{The optimized number of molecules and time slots: (a) The optimized number of molecules allocated to each transmitter node as a function of frame via STDN approach. (b) The optimized symbol durations for each transmitter node as a function of frame via DTSN approach ($\Omega = \text{4.5}\times\text{10}^{\text{-9}}~\text{m}^2/\text{s}$).  }
\end{figure*}


In Fig.~\ref{Fig:STDN_bar}, the optimized number of molecules allocated to each transmitter node is illustrated. It shows that by increasing $T_{\text{max}}$, the optimized number of molecules is increased because the BER of the system is decreased. Thus, the optimization problem (\ref{eq:STDN_A_single}) demands to increase the number of molecules for the high value of $T_{\text{max}}$. By setting $T_{\text{max}}$~=~13~ms, the number of allocated molecules to TX-1, TX-2, and TX-3 are 500, 697, and 800, respectively. In addition, in all values of $T_{\text{max}}$, TX-3 require more number of molecules than TX-1 and TX-2. It is due to the fact that TX-3 is located in a longer distance than others, therefore, the BER of it is lower than other transmitters. Then, the optimized number of molecules allocated to TX-3 is more than other transmitters. In such novel DDS, the drug release mechanism in case of utilizing STDN approach is applied by optimizing the number of molecules, i.e., the drug dosage, and uniformly allocate the releasing time between the transmitters. The release mechanism can hold Fig.~\ref{Fig:STDN_bar} to control releasing drugs into the intended location.

The optimized symbol durations as a function of $T_{\text{max}}$ for DTSN approach is illustrated in Fig.~\ref{Fig:DTSN_frame_bar}. It shows that by increasing $T_{\text{max}}$, the optimized time slots increased. Another key point to remember is that the optimized symbol duration for TX-3 is more than other transmitters from the view of the fact that the distance of TX-3 from node D is more than other transmitters. For example, by setting $T_{\text{max}}$~=~20~ms, the optimized symbol durations are $t_1$~=~3.478~ms, $t_2$~=~5.682~ms, and $t_3$~=~10.34~ms.

\begin{figure}[]
	\centering
	\scalebox{.1}{}
	\includegraphics[width=250pt]{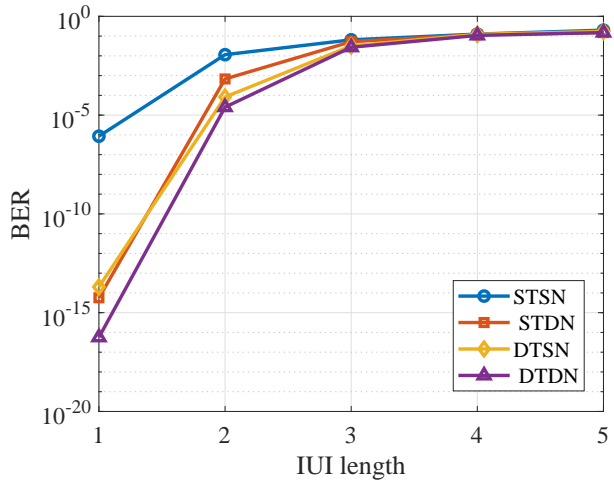}
	\caption{The minimized BER of the TDMA-based MCvD system as a function of IUI length for different approaches ($T_{\text{max}}$~=~4.5~ms and $\Omega =  \text{4.5}\times\text{10}^{\text{-9}}~\text{m}^2/\text{s}$).}
	\label{Fig:Mixture_IUI}
\end{figure}

Since the channel in MC has memory~\cite{kilinc2013receiver}, study the IUI plays a key role to identify the behavior of MCvD systems. Furthermore, in novel DDS drug releasing mechanism, it is important to release the drugs into the desired location in particular times~\cite{chahibi2013molecular}. Therefore, study the channel memory that causes the IUI effect, can improve the performance of the system. In this regard, we investigate the IUI on how to affect the BER of the system in Section~\ref{subsec:ISI_IUI}. In Fig.~\ref{Fig:Mixture_IUI}, BER of the system is illustrated as a function of IUI length by utilizing STSN, STDN, DTSN, and DTDN approaches in MODE. It is shown that the performance of the system is decreased by increasing the IUI length as a consequence of taking the previous frames' information into account. It is also shown that after IUI length of 3, the BER has no changes. Thus, we can suggest the IUI length as 3. In other words, the IUI effect for IUI length more than 3 can be neglected.


	\begin{figure*}%
	\centering
	\subfigure[]{%
		\label{Fig:iterations}%
		\includegraphics[height=150pt]{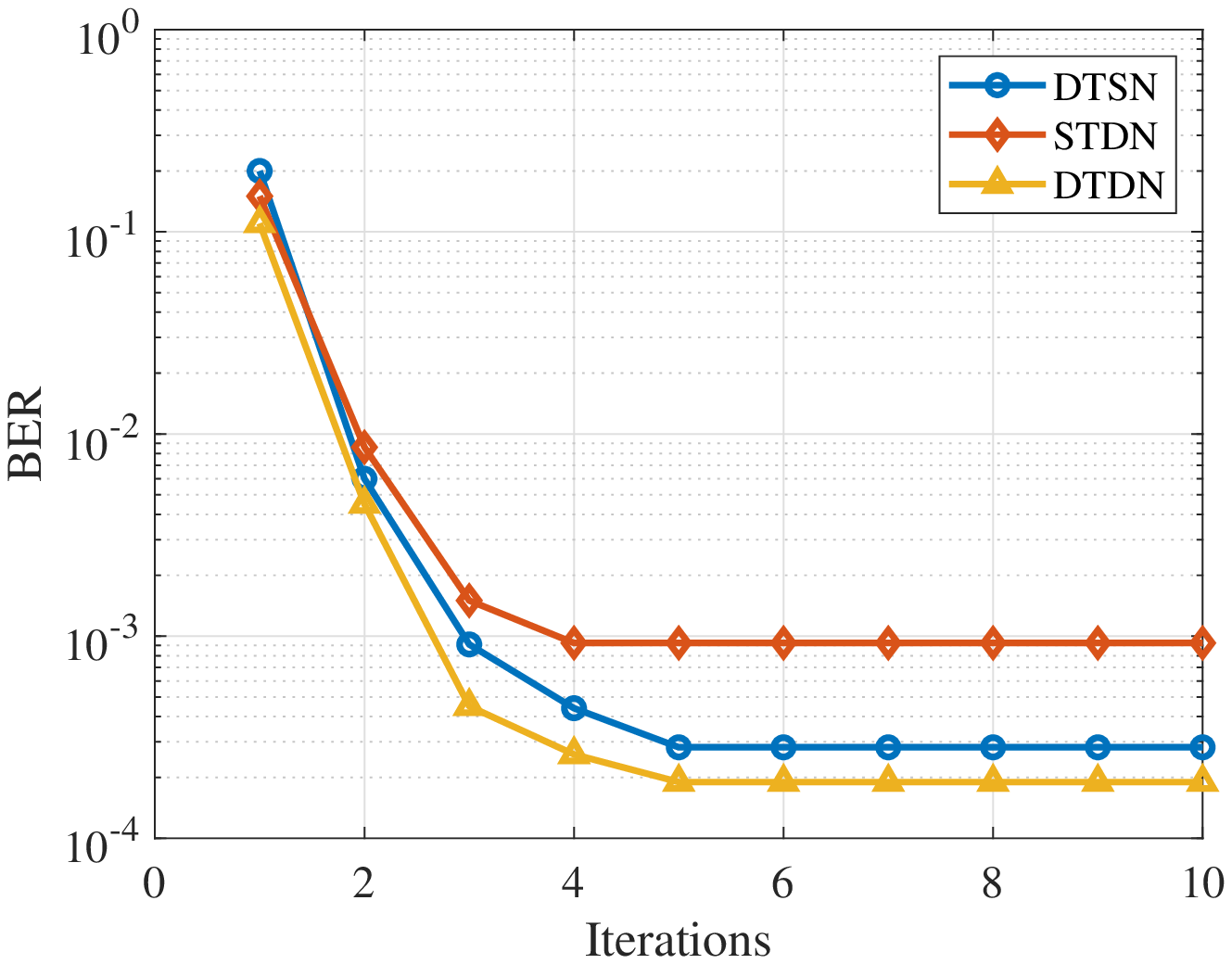}}%
	\subfigure[]{%
		\label{Fig:STDN_lower_bound}%
		\includegraphics[height=150pt]{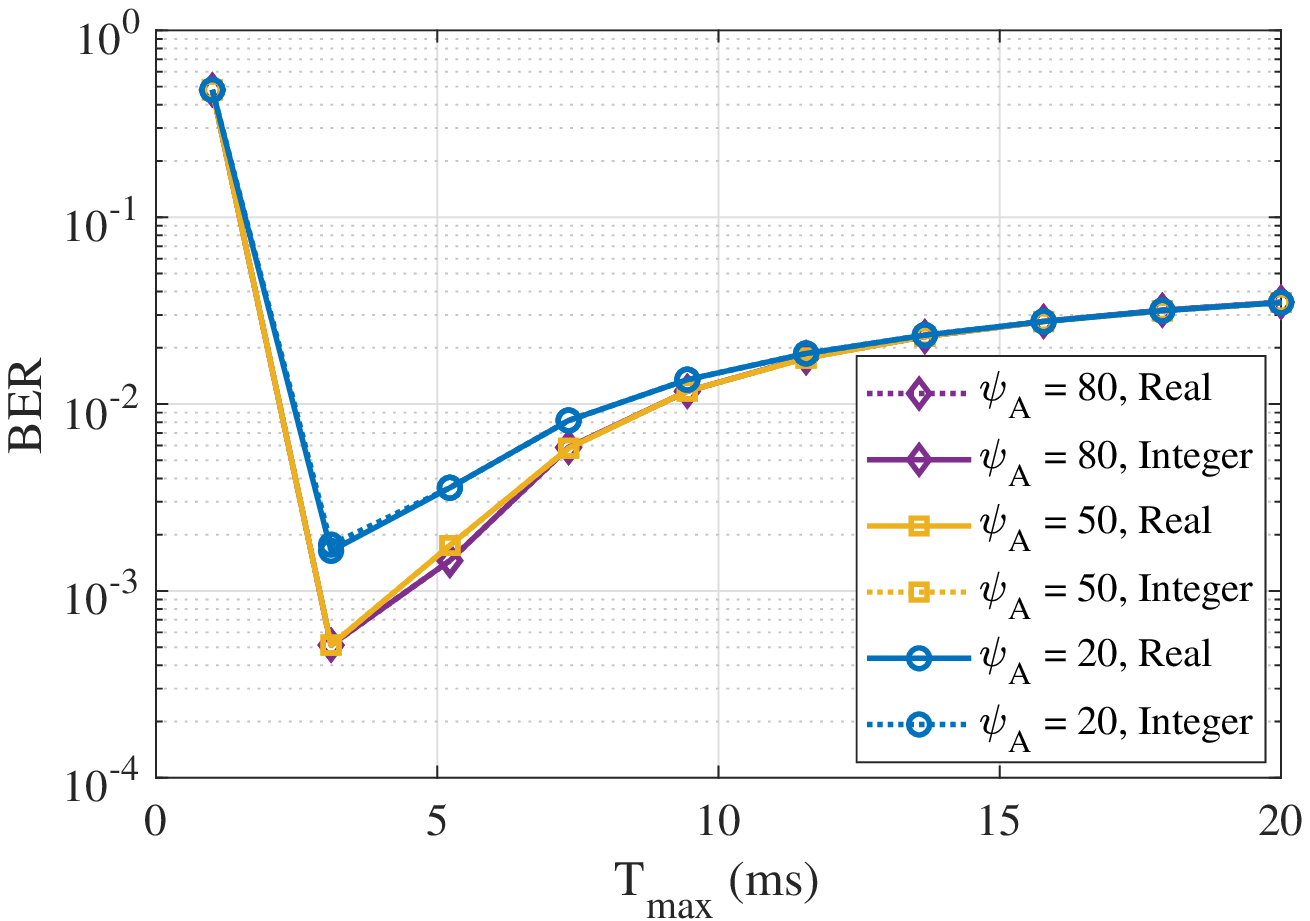}}%
	\caption{(a) The ASM convergence in DTSN, STDN, and DTDN approaches for $T_{\text{max}}=\text{10~ms}$ in TDMA-based MCvD system. (b) The BER performance of STDN approach as a function of frame for different values of $\psi_A$ in optimization problem~(\ref{eq:STDN_A_single}) by considering the number of molecules as real and integer variables ($\Psi_A = \text{800}$ and  $\Omega =  \text{4.5}\times\text{10}^{\text{-9}}~\text{m}^2/\text{s}$).}
\end{figure*}

In addition, Fig.~\ref{Fig:iterations} depicts the convergence of the alternative method proposed in (\ref{eq:DTSN_ASM}), (\ref{eq:STDN_ASM}), and (\ref{eq:ASM}). In this figure, the number of required iterations to achieve convergence for DTSN, STDN, and DTDN approaches are provided. 

In Fig.~\ref{Fig:STDN_lower_bound}, we illustrate the BER in STDN approach as a function of $T_{\text{max}}$ for different values of $\psi_A$ by considering the number of molecules as real and integer variables. For $\psi_A = \text{20}$, the difference between BER for integer and real $\boldsymbol{A}$ is $\text{1.5}\times \text{10}^{\text{-4}}$. However, for $\psi_A > \text{80}$, the BER difference is smaller than $\text{4}\times \text{10}^{\text{-7}}$ because increasing $\psi_A$ causes decreasing the BER difference. Therefore, considering the number of molecules as a real variable does not affect extremely the BER of the system.

\section{Conclusion and Future Works} \label{sec:conclusion}
In this paper, the drug release management is studied in case of utilizing dynamic TDMA-based MCvD system. In this regard, the number of molecules, e.g., the drug dosage in such novel DDS, allocated to each transmitter node and symbol durations are assessed. Furthermore, four approaches are introduced in three scenarios namely MDE, MODE, and SDE. At first, static state of symbol durations and number of molecules which are uniformly distributed between the transmitters, as STSN, is evaluated. Next, the other approaches as DTSN and STDN are considered, which in the first one, the symbol durations are optimized by allocating an equal number of molecules to each transmitter node. The other one is based on optimizing the number of molecules allocated to each transmitter node and identical symbol durations. The last approach is DTDN where both symbol durations and the number of molecules are optimized simultaneously. The aforementioned scenarios are evaluated in terms of the type of information molecules and the medium, i.e., the diffusion coefficient. Moreover, the TDMA-based MCvD system is investigated in terms of the mean and the variance of the received molecules which are contained interference. We also study the error probability of the received bits through mathematical manipulations. Since our optimization problems are MOP, we investigate the optimization solutions for them. The scalarization method by employing WSM is studied as the solution of the proposed optimization problems. At the numerical analysis, we evaluate the scenarios for drug releasing management in connection with the complexity of the nanomachines and the time required to release the drug to the desired location. Furthermore, we presented the performance of optimization the symbol durations and number of molecules. The effect of IUI is also investigated. We solved the optimization problems by utilizing ASM and the convergences of the proposed solutions had been assessed.

As the future work, one can consider the mobile transmitter/receiver nodes in TDMA-based MCvD system and investigate the STSN, STDN, DTSN, and DTDN approaches in order to impact the performance of such novel DDS. Furthermore, interested readers can study the learning methods, e.g., reinforcement learning, to design the mechanism of drug releasing in applications of MC such as HIV, genetic diseases, and cancer therapy.
\appendices
\section{The Calculations of Mean and Variance for IUI} \label{Appendix I}
In this appendix, we aim to express the calculation of mean and variance of IUI. As we discussed, the molecules that are leaked into the current time slot distributed normally. The distribution of molecules that are leaked from the previous frame $u$ and TX-$j$ into current frame $n$ for transmitter TX-$s$ is given by:
\begin{align}
\begin{split} \label{eq:IUI_first_app}
M^s_{\text{IUI}} [n,u,j] =&  \ \mathcal{N}\bigg(A_j \ x_j[n - u] Y^u_{j,s}, A_j \ x_j[n - u] Y^u_{j,s} \big(1 - Y^u_{j,s} \big)\bigg).
\end{split}
\end{align}
Furthermore, the distribution of molecules that stand at IUI effect in current frame $n$ from TX-$j$ into TX-$s$ is
\begin{align}
\begin{split} \label{eq:IUI_first_app_2}
M^s_{\text{IUI-C}} [n,j] =&  \ \mathcal{N}\bigg(A_j \ x_j[n] H_{j,s}, A_j \ x_j[n]  H_{j,s} \big(1 - H_{j,s} \big) \bigg).
\end{split}
\end{align}


The mean of $M^s_{\text{IUI}} [n,u,j]$ is calculated as:
\begin{align}
\begin{split} \label{eq:IUI_second_app}
E(M^s_{\text{IUI}} [n,u,j]) = & \ 0.5 \ \bigg( E(M^s_{\text{IUI}} [n,u,j] \mid x_j[n - u] =0) \ +\  E(M^s_{\text{IUI}} [n,u,j] \mid x_j[n - u] =1) \bigg) \\=& \  0.5 \  A_j \ Y^u_{j,s},
\end{split}
\end{align}
where $0.5$ is the probability of sending bit ``1'' and ``0''. In addition, the mean of $M^s_{\text{IUI-C}} [n,q]$ is
\begin{align}
\begin{split} \label{eq:IUI_second_app_2}
E(M^s_{\text{IUI-C}} [n,j]) = & \ 0.5 \ \bigg( E(M^s_{\text{IUI-C}} [n,j] \mid x_j[n] =0) \ +\  E(M^s_{\text{IUI-C}} [n,j] \mid x_j[n] =1) \bigg) \\=& \  0.5 \  A_j \ H_{j,s}.
\end{split}
\end{align}

The variance of $M^s_{\text{IUI}} [n,u,j]$ is given by:
 \begin{align}
 \begin{split}
 Var(M^s_{\text{IUI}} [n,u,j]) = & \ E((M^s_{\text{IUI}} [n,u,j])^2) - E^2(M^s_{\text{IUI}} [n,u,j]), \label{eq:IUI_third_app}
 \end{split}
 \end{align}
 and the variance of $M^s_{\text{IUI-C}} [n,j]$ is calculated as:
  \begin{align}
 \begin{split}
 Var(M^s_{\text{IUI-C}} [n,j]) = & \ E((M^s_{\text{IUI-C}} [n,j])^2) - E^2(M^s_{\text{IUI-C}} [n,j]), \label{eq:IUI_third_app_2}
 \end{split}
 \end{align}
In the following, we derive $E((M^s_{\text{IUI}} [n,u,j])^2)$ as
\begin{align}
\begin{split} \label{eq:IUI_app}
E((M^s_{\text{IUI}} [n,u,j])^2) = & \ 0.5 \ \bigg(E\big((M^s_{\text{IUI}} [n,u,j])^2 \mid x_j[n - u]=0\big) + E\big((M^s_{\text{IUI}} [n,u,j])^2 \mid x_j[n - u]=1\big) \bigg) \\= & \ 0.5  A_j \ Y^u_{j,s} \big(1 - Y^u_{j,s} \big) - (A_j Y^u_{j,s})^2. 
\end{split}
\end{align}
Consequently, $E((M^s_{\text{IUI-C}} [n,j])^2)$ is 
\begin{align}
\begin{split} \label{eq:IUI_app_2}
E((M^s_{\text{IUI-C}} [n,j])^2) = & \ 0.5 \ \bigg(E\big((M^s_{\text{IUI-C}} [n,j])^2 \mid x_j[n]=0\big) + E\big((M^s_{\text{IUI-C}} [n,j])^2 \mid x_j[n]=1\big) \bigg) \\= & \ 0.5  A_j  H_{j,s} \big(1 - H_{j,s} \big) - (A_j H_{j,s})^2. 
\end{split}
\end{align}

At last, the variance of the molecules released from frame $u$ and TX-$j$ but received in the current frame $n$ for TX-$s$ is calculated by (\ref{eq:IUI_second_app}), (\ref{eq:IUI_third_app}), and (\ref{eq:IUI_app}) as 
\begin{align}
\begin{split}
Var(M^s_{\text{IUI}} [n,u,j]) = & \ 0.5 A_j Y^u_{j,s} - A_j (Y^u_{j,s})^2 \big( 0.5 - 1.25 A_j \big), \label{eq:var_app}
\end{split}
\end{align}
and the variance of the molecules which are released in current frame from TX-$j$ but received in TX-$s$' time slot is derived by (\ref{eq:IUI_second_app_2}), (\ref{eq:IUI_third_app_2}), and (\ref{eq:IUI_app_2}) as
\begin{align}
\begin{split}\label{eq:var_app_2}
Var(M^s_{\text{IUI-C}} [n,j]) = & \ 0.5 A_j H_{j,s} - A_j (H_{j,s})^2 \big( 0.5 - 1.25 A_j \big). 
\end{split}
\end{align}

Finally, the mean and variance for IUI is given as:
\begin{subequations} \label{eq:IUI_calculatation}
\begin{align}
 E(M^s_{\text{IUI}} [n]) =& \sum_{u = 1}^{U} \sum_{j=1}^{r} E(M^s_{\text{IUI}} [n,u,j]) + \sum_{j=1}^{s-1} E(M^s_{\text{IUI-C}} [n,j])\\
 Var(M^s_{\text{IUI}} [n]) =& \sum_{u = 1}^{U} \sum_{j=1}^{r} Var(M^s_{\text{IUI}} [n,u,j]) + \sum_{j=1}^{s-1} Var(M^s_{\text{IUI-C}} [n,j]).
\end{align}
\end{subequations}

\section{Proof The Convexity of The Optimization Problem (\ref{eq:DTSN_t_single}) }  \label{Appendix III}
In this appendix, the convexity of the objective function in (\ref{eq:DTSN_t_single}) is proved. The first derivative of (\ref{eq:probability}) with respect to $t_s$ is calculated as
\begin{align}
\begin{split}
\dfrac{\partial}{\partial t_s} P^s_{e}[n] = \  &\dfrac{\partial P^s_{e}[n]}{\partial \mu_{0_s}} \ \dfrac{\partial \mu_{0_s}}{d t_s} + \dfrac{\partial P^s_{e}[n]}{\partial \mu_{1_s}} \ \dfrac{\partial \mu_{1_s}}{\partial t_s} +\dfrac{\partial P^s_{e}[n]}{\partial \sigma^2_{0_s}} \ \dfrac{\partial \sigma^2_{0_s}}{\partial t_s} + \dfrac{\partial P^s_{e}[n]}{\partial \sigma^2_{1_s}} \ \dfrac{\partial \sigma^2_{1_s}}{\partial t_s}. \label{app4:first_der}
\end{split}
\end{align}

The first derivative of (\ref{eq:probability}) with respect to the mean and variances are given as:
\begin{subequations} \label{app3_mean_var_der}
	\begin{align}
	\begin{split}
	\dfrac{\partial P^s_{e}[n]}{\partial \mu_{0_s}} =& \ \frac{1}{4 \sqrt{2 \pi}} \bigg( \frac{1}{\sqrt{2 \sigma^2_{0_s}}} \text{exp} \big( - \frac{\mu^2_{0_s}}{2 \sigma^2_{0_s}} \big) \bigg), \label{app3:respect_mu0}
	\end{split}
	\\
	\begin{split}
	\dfrac{\partial P^s_{e}[n]}{\partial \mu_{1_s}} =& \ \frac{-1}{4 \sqrt{2 \pi}} \bigg( \frac{1}{\sqrt{2 \sigma^2_{1_s}}} \text{exp} \big( - \frac{\mu^2_{1_s}}{2 \sigma^2_{1_s}} \big) \bigg), \label{app3:respect_mu1}
	\end{split}
	\\
	\begin{split}
	\dfrac{\partial P^s_{e}[n]}{\partial \sigma^2_{0_s}} =& \ \frac{1}{8 \sqrt{2 \pi}} \bigg(  (\tau_s - \mu_{0_s}) (\sigma^2_{0_s})^{-\frac{3}{2}} \text{exp} \big( \dfrac{- (\tau_s - \mu_{0_s})^2 }{2 \sigma^2_{0_s}} \big)  \bigg), \label{app3:respect_var0}
	\end{split}
	\\
	\begin{split}
	\dfrac{\partial P^s_{e}[n]}{\partial \sigma^2_{1_s}} =& \ \frac{-1}{8 \sqrt{2 \pi}} \bigg(  (\tau_s - \mu_{1_s}) (\sigma^2_{1_s})^{-\frac{3}{2}} \text{exp} \big( \dfrac{- (\tau_s - \mu_{1_s})^2 }{2 \sigma^2_{1_s}} \big)  \bigg). \label{app3:respect_var1}
	\end{split}
	\end{align}
\end{subequations}
The first derivative of the mean and the variance of the number of received molecules with respect to $t_s$ are derived as follows:
\begin{subequations} \label{eq:der_T}
	\begin{align}
		\begin{split}
		\dfrac{\partial \mu_{0_s}}{\partial t_s} =& \ 0.5 g(t_s) \bigg( \sum_{j=1}^{r} A_j  + \sum_{j=1}^{s-1} A_j \bigg), \label{eq:der_T_mu0}
		\end{split}
		\\
		\begin{split}
		\dfrac{\partial \mu_{1_s}}{\partial t_s} =& \ A_s g(t_s) + \dfrac{\partial \mu_{0_s}}{\partial t_s}, \label{eq:der_T_mu1}
		\end{split}
		\\
		\begin{split} \label{eq:der_T_var0}
		\dfrac{\partial \sigma^2_{0_s}}{\partial t_s}=& \ 0.5 \sum_{u = 1}^{U} \sum_{j=1}^{r} \big\{A_j g(\lambda^u_{j,s}) \big( 1 - Y^u_{j,s} (1 - 2.5 A_j) \big)\big\} + 0.5 g(t_s) \sum_{j=1}^{s-1} \big\{  A_j \big( 1 - H^u_{j,s} (1 - 2.5 A_j) \big) \big\},
		\end{split}
		\\
		\begin{split}
		\dfrac{\partial \sigma^2_{1_s}}{\partial t_s}=& \ A_s g(t_s) \big(1 - 2 g(t_s) P_h(\mathbf{X_s},t_s)\big) + \dfrac{\partial \sigma^2_{0_s}}{\partial t_s}, \label{eq:der_T_var1}
		\end{split}
	\end{align}
\end{subequations}
where $g (t_s) $ is the first derivative of (\ref{eq:CDF}) with respect to $t_s$ which is given by
\begin{align}
	\begin{split}
	g (t_s) =& \int_{-d}^{d} \int_{-\sqrt{d^2 - z^2}}^{\sqrt{d^2 - z^2}} \int_{-\sqrt{d^2 - z^2 - y^2}}^{\sqrt{d^2 - z^2 - y^2}} \bigg(   \dfrac{\Omega^2 t_s \big( (z+D_z)^2 - u_z^2 t_s^2 + (y+D_y)^2 - u_y^2 t_s^2 + (x+D_x)^2}{32 \pi^{3/2} (\Omega^3 t^3_s)^{3/2}} \\& - \dfrac{u_x^2 t_s^2 - 6 \Omega t_s \big)}{32 \pi^{3/2} (\Omega^3 t^3_s)^{3/2}} \bigg)  \exp\bigg( \dfrac{(z+D_z)^2 - 2u_z t_s(z+D_z) + u_z^2 t_s^2 + (y+D_y)^2}{4 \Omega t_s} \\&- \dfrac {2u_y t_s(y+D_y) + u_y^2 t_s^2 + (x+D_x)^2 - 2 u_x t_s(y+D_y) + u_x^2 t_s^2}{4 \Omega t_s}  \bigg) 
	~\text{d}x~\text{d}y~\text{d}z. \label{eq:der_CDF}
	\end{split}
\end{align}

The second derivative of the mean and the variance of the number of received molecules is required to calculate the second derivative of (\ref{eq:probability}) with respect to $t_s$. The second derivative of the mean and the variance of the received molecules are given by:

\begin{subequations} \label{eq:der_T_sec}
	\begin{align}
	\begin{split}
	\dfrac{\partial^2 \mu_{0_s}}{\partial t_s^2} =& \ 0.5 g'(t_s) \bigg( \sum_{j=1}^{r} A_j  + \sum_{j=1}^{s-1} A_j \bigg), \label{eq:der_T_mu0_sec}
	\end{split}
	\\
	\begin{split}
	\dfrac{\partial^2 \mu_{1_s}}{\partial t_s^2} =& \ A_s g'(t_s) + \dfrac{\partial^2 \mu_{0_s}}{\partial t_s^2}, \label{eq:der_T_mu1_sec}
	\end{split}
	\\
	\begin{split} \label{eq:der_T_var0_sec}
	\dfrac{\partial^2 \sigma^2_{0_s}}{\partial t_s^2}=& \ 0.5 \sum_{u = 1}^{U} \sum_{j=1}^{r} \bigg\{A_j g'(t_s) - A_jg'(t_s) Y^u_{j,s} (1 - 2.5 A_j) - A_j g(t_s) g(\lambda^u_{j,s}) (1 - 2.5 A_j) \bigg\} \\ &- 0.5 g(t_s) g'(t_s) \sum_{j=1}^{s-1}  (1 - 2.5 A_j),
	\end{split}
	\\
	\begin{split}
	\dfrac{\partial^2 \sigma^2_{1_s}}{\partial t_s^2}=& \ A_s \bigg( g'(t_s) - 4g(t_s) g'(t_s) P_h(\mathbf{X_s},t_s) - 2g^2 (t_s) \bigg) + \dfrac{\partial^2 \sigma^2_{0_s}}{\partial t_s^2}, \label{eq:der_T_var1_sec}
	\end{split}
	\end{align}
\end{subequations}
where $g'(t_s)$ is the second derivative of (\ref{eq:CDF}) with respect to $t_s$. After some manipulations and by given that $\mu_{1_s} > \mu_{0_s}$ and $\sigma^2_{1_s} > \sigma^2_{0_s}$, the second derivative of BER with respect to $t_s$ is positive. 
Finally, due to the fact that the objective function in (\ref{eq:DTSN_t_single}) on the convex set $t_s$ is convex, we conclude that the optimization problem (\ref{eq:DTSN_t_single}) is convex on $t_s$.

\section{Proof The Convexity of The Optimization Problem (\ref{eq:STDN_A_single})}  \label{Appendix IV}
In this appendix, we provide the proof of the convexity of error probability function in (\ref{eq:probability}). The first derivative of (\ref{eq:probability}) with respect to $A_s$ is equal to
\begin{align}
\begin{split}
\dfrac{\partial}{\partial A_s} P^s_{e}[n] = \  &\dfrac{\partial P^s_{e}[n]}{\partial \mu_{0_s}} \ \dfrac{\partial \mu_{0_s}}{d A_s} + \dfrac{\partial P^s_{e}[n]}{\partial \mu_{1_s}} \ \dfrac{\partial \mu_{1_s}}{\partial A_s} +\dfrac{\partial P^s_{e}[n]}{\partial \sigma^2_{0_s}} \ \dfrac{\partial \sigma^2_{0_s}}{\partial A_s} + \dfrac{\partial P^s_{e}[n]}{\partial \sigma^2_{1_s}} \ \dfrac{\partial \sigma^2_{1_s}}{\partial A_s}. \label{app3:first_der}
\end{split}
\end{align}

The first derivative of (\ref{eq:probability}) with respect to the mean and variances are calculated in (\ref{app3_mean_var_der}).

 The first derivative of mean and variances with respect to $A_s$ are derived as below:
 
\begin{subequations}
 \begin{align}
 	\begin{split}
 	 \dfrac{\partial \mu_{0_s}}{\partial A_s} =& \ 0.5 \sum_{u=1}^{U} Y^u_{s,s}, \label{app3:mu0_respect}
 	\end{split}
 	\\
 \begin{split}
\dfrac{\partial \mu_{1_s}}{\partial A_s} =& \ P_h (\mathbf{X_s},t_s) + 0.5 \sum_{u=1}^{U} Y^u_{s,s}, \label{app3:mu1_respect}
 \end{split}
 \\
 \begin{split}
\dfrac{\partial \sigma^2_{0_s}}{\partial A_s} =& \ 0.5 \sum_{u=1}^{U} \big\{ Y^u_{s,s} - (Y^u_{s,s})^2 \big( 0.5 - 2.5 A_s \big)\big\}, \label{app3:var0_respect}
 \end{split}
 \\
 \begin{split}
 \dfrac{\partial \sigma^2_{1_s}}{\partial A_s} =& \ P_h (\mathbf{X_s},t_s) \big(1 - P_h (\mathbf{X_s},t_s) \big) + 0.5 \sum_{u=1}^{U} \big\{ Y^u_{s,s} - (Y^u_{s,s})^2 \big( 0.5 - 2.5 A_s \big)\big\}. \label{app3:var1_respect}
 \end{split}
 \end{align}
\end{subequations}

The second derivative of (\ref{eq:probability}) with respect to $A_s$ is
 \begin{align}
 	\begin{split}
\dfrac{\partial^2}{\partial A^2_s} P^s_{e}[n] =& \ 1.25  \sum_{u=1}^{U} (Y^u_{s,s})^2  \times \bigg( \dfrac{\partial P^s_{e}[n]}{\partial \sigma^2_{0_s}} +  \dfrac{\partial P^s_{e}[n]}{\partial \sigma^2_{1_s}} \bigg). \label{app3:sec_der}
\end{split}
 \end{align}
 
 By considering (\ref{app3:sec_der}), the values of second derivative of (\ref{eq:probability}) with respect to $A_s$ is positive, if the following condition is satisfied
 \begin{align}
 	\mu_{0_s} < \tau_s < \mu_{1_s}.
 \end{align}
 By employing ML as the detection technique, this condition is always right. Since (\ref{eq:STDN_A_single}) is a summation on $P^s_{e}$, therefore, (\ref{app3:sec_der}) is positive.
 Finally, objective function in (\ref{eq:STDN_A_single}) on the convex set $A_s$ is convex, therefore, the optimization problem (\ref{eq:STDN_A_single}) is convex on $A_s$.

\bibliographystyle{ieeetr}
\bibliography{Hamid_Khoshfekr}
\end{document}